\documentclass[twocolumn]{aastex631}
\usepackage{booktabs, hyperref, comment, tabularx}
\usepackage[T1]{fontenc}

\begin{document}

\title{Discovery and Detailed Study of the M31 Classical Nova AT~2023tkw: Evidence for Internal Shocks}

\shorttitle{M31 Nova: AT 2023tkw }
\shortauthors{Basu et al.}

\correspondingauthor{Judhajeet Basu}
\email{judhajeet.basu@iiap.res.in}

\author[0000-0001-7570-545X]{Judhajeet Basu}
\affiliation{Indian Institute of Astrophysics, II Block, Koramangala, Bengaluru 560034, India}
\affiliation{Pondicherry University, R.V. Nagar, Kalapet, Pondicherry 605014, India}

\author[0009-0008-6428-7668]{Ravi Kumar}
\affiliation{Department of Aerospace Engineering, Indian Institute of Technology Bombay, Powai, Mumbai 400076, India}

\author[0000-0003-3533-7183]{G.C. Anupama}
\affiliation{Indian Institute of Astrophysics, II Block, Koramangala, Bengaluru 560034, India}

\author[0000-0002-3927-5402]{Sudhanshu Barway}
\affiliation{Indian Institute of Astrophysics, II Block, Koramangala, Bengaluru 560034, India}

\author[0000-0002-8077-5572]{Peter H. Hauschildt}
\affiliation{Hamburger Sternwarte, University of Hamburg, Gojenbergsweg 112, 21029 Hamburg, Germany}

\author[0009-0000-5909-293X]{Shatakshi Chamoli}
\affiliation{Indian Institute of Astrophysics, II Block, Koramangala, Bengaluru 560034, India}
\affiliation{Pondicherry University, R.V. Nagar, Kalapet, Pondicherry 605014, India}

\author[0000-0002-7942-8477]{Vishwajeet Swain}
\affiliation{Department of Physics, Indian Institute of Technology Bombay, Powai, Mumbai 400076, India}

\author[0000-0002-6112-7609]{Varun Bhalero}
\affiliation{Department of Physics, Indian Institute of Technology Bombay, Powai, Mumbai 400076, India}

\author[0000-0003-2758-159X]{Viraj Karambelkar}
\affiliation{Cahill Center for Astrophysics, California Institute of Technology, 1200 E. California Blvd. Pasadena, CA 91125, USA}

\author[0000-0002-5619-4938]{Mansi Kasliwal}
\affiliation{Cahill Center for Astrophysics, California Institute of Technology, 1200 E. California Blvd. Pasadena, CA 91125, USA}

\author[0000-0001-8372-997X]{Kaustav K. Das}
\affiliation{Cahill Center for Astrophysics, California Institute of Technology, 1200 E. California Blvd. Pasadena, CA 91125, USA}

\author[0000-0002-8977-1498]{Igor Andreoni}
\affiliation{Department of Physics and Astronomy, University of North Carolina at Chapel Hill, Chapel Hill, NC 27599-3255, USA}

\author[0000-0003-2091-622X]{Avinash Singh}
\affiliation{Department of Astronomy, The Oskar Klein Center, Stockholm University, AlbaNova University Center, SE 106 91 Stockholm, Sweden}

\author[0000-0002-0525-0872]{Rishabh Singh Teja}
\affiliation{Indian Institute of Astrophysics, II Block, Koramangala, Bengaluru 560034, India}
\affiliation{Pondicherry University, R.V. Nagar, Kalapet, Pondicherry 605014, India}

\begin{abstract}
We present a detailed analysis of an extragalactic slow classical nova in M31 exhibiting multiple peaks in its light curve. Spectroscopic and photometric observations were used to investigate the underlying physical processes. Shock-induced heating events resulting in the expansion and contraction of the photosphere are likely responsible for the observed multiple peaks. Deviation of the observed spectrum at the peak from the models also suggests the presence of shocks. The successive peaks occurring at increasing intervals could be due to the series of internal shocks generated near or within the photosphere. Spectral modeling suggests a low-mass white dwarf accreting slowly from a companion star. The ejecta mass, estimated from spectral analysis, is $\sim 10^{-4}\mathrm{M_{\odot}}$, which is typical for a slow nova. We estimate the binary, by comparing the archival HST data and eruption properties with stellar and novae models, to comprise a 0.65 $\mathrm{M_{\odot}}$ primary white dwarf and a K III cool evolved secondary star.

\end{abstract}

\keywords{novae, cataclysmic variables --- shock waves --- galaxies: individual (M31) --- techniques: image processing, photometric, spectroscopic
}


\section{Introduction} \label{sec:intro}
Novae are cataclysmic events caused by a thermonuclear runaway reaction on the surface of a white dwarf (WD), accreting matter from a secondary star. The consequences are the brightening of the binary system by several orders of magnitudes accompanied by ejection of matter  \citep{STARRFIELD1999371,2008clno.book.....B,Starrfield_2016, Kato_2022}.

Nova systems typically host a dwarf secondary star \citep{Morales_2021}. However, giant secondaries are not rare \citep{Schaefer_2021, Chomiuk_2021}. Systems with low WD mass ($\lesssim 1~\mathrm{M_{\odot}}$) and/or low accretion rates ($\lesssim 10^{-9}~\mathrm{M_{\odot} yr^{-1}}$) take a long time, more than 100~yrs, to accumulate H-rich material necessary for the thermonuclear runaway reaction. Such novae have been observed to erupt only once and are termed classical novae (CNe). Meanwhile, some systems with high WD mass ($\gtrsim 1.2~\mathrm{M_{\odot}}$) and high accretion rates ($10^{-6}-10^{-8}~\mathrm{M_{\odot} yr^{-1}}$) need less ignition mass to erupt and thus are seen frequently in outbursts and are termed recurrent novae (RNe) \citep{Yaron_2005, Shara_2018}.

Novae show diversity in their light curves and spectra, leading to classifications based on their speed class (very fast, fast, moderate, slow, and very slow; \citealt{Warner_2003}), spectral class (\ion{Fe}{2}, He/N, and hybrid; \citealt{Williams_2012}), and light curve morphology (smooth, cuspy, plateau, flat-topped, jittery, dusty, and oscillatory; \citealt{strope_2010}). Different spectral phases are thought to originate at different evolutionary stages of a nova depending on the opacity, density, and ionization of the ejecta. Systems that evolve rapidly can cause some of these spectral phases to be missed, particularly during the early evolution \citep{Aydi_2024}. However, rapid responses in modern times have enabled observations during most of the spectral phases, even in the fast recurrent novae (U Sco: \citealt{2010MNRAS.408L..71B}, \citealt{Anupama_2013}; RS Oph: \citealt{Pandey_2022}). Models have suggested that the WD's mass ($\mathrm{M_{WD}}$), accretion rate ($\mathrm{\dot M}$), and the intrinsic temperature ($\mathrm{T_{WD}}$) or the luminosity ($\mathrm{L_{WD}}$) are the three most important parameters responsible for describing the novae behavior in a cataclysmic variable (CV) system \citep{Yaron_2005, Starrfield_2012, Hillman_2016}. Other secondary parameters affecting the CV properties include the binary separation, composition of the accreted matter, the type of secondary star, and magnetic field strength of the WD.

Recent advances in multi-wavelength observations of novae have revealed $\gamma$-ray and hard X-ray emissions concurrent with optical emission (\citealt{Chomiuk_2021} and references therein). The presence of internal shocks in novae has become more evident and is also backed up by models \citep{Metzger_2014, Martin_2018}. However, these high-energy radiations, constrained by sensitivity, can be tested only on galactic novae. For extra-galactic novae, we must rely on the effect of shocks seen in UV-optical bands.

M31, the nearby large galaxy, is a treasure trove for nova hunters, given its suitable distance, size, inclination, visibility, and high nova rates \citep{Darnley_2006, Rector_2022}. Such outstanding possibilities have led to astronomers surveying M31 for decades. More than 1100 novae have been discovered in M31 in the past century \citep{Pietsch_2010, Darnley_2020}. M31 has been home to numerous recurrent novae, almost double than discovered in our Galaxy \citep{Shafter_2015}. The most exotic ones have a recurrence period of $\lesssim$ 10 years \citep{Shafter_2024}, including the unusual RN with a one-year period M31N 2008-12a \citep{basu_2024}. These systems have massive WDs and thus are one of the nearest type Ia supernova prospects through the single degenerate scenario \citep{Kato_2014}. These objects, along with a diverse population of novae in M31, establish the motivation to survey M31. 

This paper discusses a slow classical nova AT~2023tkw, discovered by the fully robotic 0.7~m GROWTH-India Telescope (GIT; \citealt{harsh_2022a}) during its M31 survey. In \S \ref{sec:discovery}, we talk about the discovery, classification, and follow-up observations of the object. We discuss the imaging and spectroscopic analysis in \S \ref{sec:lc_evol} and \S \ref{sec:spec_evol}. \S \ref{sec:progenitor} sheds light on the binary components of this system. We end with a discussion on the properties of the nova.


\section{Discovery and  Observations}
\label{sec:discovery}

M31 is being observed at a daily cadence by the GIT in the  SDSS $g^{\prime}$ and $r^{\prime}$ bands in the months of July to February every year starting from 2022 under a proposal to search for nova-like transients through image subtraction techniques using reference images from Pan-STARRS of each field daily, following the methods in \cite{harsh_2022b}. Candidates were screened via our vetting interface. On 19th September 2023, we detected an unidentified source at just over 20 mags in both $g^{\prime}$ and $r^{\prime}$ bands at the J2000 coordinates (RA, Dec) = (00:41:24.169, +41:08:05.26). Checks for minor planets using \texttt{mpchecker}\footnote{\href{https://minorplanetcenter.net/cgi-bin/checkmp.cgi}{https://minorplanetcenter.net/cgi-bin/checkmp.cgi}} showed no matches. The transient was not reported to TNS earlier. No crossmatch was found either in the M31 nova catalog\footnote{\href{https://www.mpe.mpg.de/~m31novae/opt/m31/M31_table.html}{https://www.mpe.mpg.de/$\sim$m31novae/opt/m31/M31\_table.html}}, or in the foreground galactic variable star search\footnote{\href{https://www.aavso.org/vsx/index.php}{https://www.aavso.org/vsx/index.php}} or in \texttt{SIMBAD}\footnote{\href{https://simbad.cds.unistra.fr/simbad/}{https://simbad.cds.unistra.fr/simbad/}}. The new source was named GIT20230919aa and reported to the Transient Name Server \citep{ravi_2023}, which designated it as AT 2023tkw. We also note its detection by the Zwicky Transient Facility (ZTF; \citealt{Bellm_2019, Dekany_2020, Graham_2019}) under the name ZTF23aayatam. The object was found to be brightening, i.e. it was discovered before it attained its peak magnitude. Spectroscopic observations followed once it was bright enough, which led to its classification as a nova in its \ion{Fe}{2} phase \citep{Basu_2023}.

\subsection{Photometry}
Optical photometric observations of AT 2023tkw were performed using the robotic GIT. We supplement these observations with forced photometry obtained using the ZTF Observing System, installed on the 48-inch Samuel Oschin Telescope (Schmidt-type) at the Palomar Observatory.

\subsubsection{GROWTH-India Telescope}

The GIT, located at the Indian Astronomical Observatory (IAO), Hanle, equipped with a 0.7$^{\circ}$ field of view, was used to obtain images of the object in the Sloan $g^{\prime},r^{\prime},i^{\prime}$ and $z^{\prime}$ filters. After the discovery on 19 September 2023, we performed dedicated observations of the field starting from 20 September till 25 December 2023, with a roughly daily cadence. Images were acquired in single or multiple co-added 300~s exposures, depending on the brightness of the object, sky conditions, and filter used. The images were processed on a nightly basis following the steps mentioned in \cite{harsh_2022a}. Host-subtracted photometry was obtained after performing image subtraction with Pan-STARRS DR1 \citep{chambers_2016} reference images as described in \cite{harsh_2022b}.

\subsubsection{Zwicky Transient Facility}
The ZTF forced-photometry service \citep{ztffp} was used to obtain photometry of the object in the ZTF-$g$ and ZTF-$r$ filters, following a roughly 3-day cadence. These observations supplement the dense light curve obtained with the GIT. The ZTF light curve revealed a long pre-discovery plateau of the object starting from $19^{\text{th}}$ July 2023 till its first rise $\sim$2 months later. The photometry table obtained from the service was carefully filtered following the steps described in \cite{masci2023newforcedphotometryservice}.

\subsection{Spectroscopy}
The spectra of the object were taken in optical and infrared (IR) using the telescopes and instruments given in Table~\ref{tab:spec_obs}. All the optical spectra were calibrated with the nearest broadband $g^{\prime},r^{\prime},i^{\prime}$ magnitudes owing to the dense photometric coverage with the GIT. The extrapolated $z^{\prime}$ mag was additionally used to calibrate the Keck/LRIS spectrum. The velocity measured from the full width at half maxima (FWHM) of emission lines was deconvolved with the instrument's resolution before being reported in the text.
\label{subsec:spec}
\begin{table}[h!]
    \centering
    \begin{tabular}{ccccc}
    \toprule
        Date of Obs & Tel/Inst & R & Exp & Airmass\\
         (UT)       &          &        & (s)  &  \\
         \midrule
         2023-10-05 07:46 & P200/DBSP     & 1000 & 3784 & 1.01 \\
         2023-10-18 19:33 & HCT/HFOSC     & 600  & 3600 & 1.10 \\
         2023-10-22 08:19 & Keck I/LRIS   & 750  & 300 & 1.09 \\
         2023-10-26 13:48 & HCT/HFOSC     & 600  & 3600 & 1.31 \\
         2023-10-26 14:54 & HCT/HFOSC     & 600  & 3100 & 1.13 \\
         2023-10-29 06:46 & Keck II/NIRES & 1000 & 300  & 1.17 \\
         2023-11-19 13:24 & HCT/HFOSC     & 600  & 3600 & 1.12 \\
     \bottomrule
    \end{tabular}
    \caption{Log of spectroscopic observations}
    \label{tab:spec_obs}
\end{table}

\subsubsection{P200/DBSP}
The Double Beam Spectrograph (DBSP) is a low resolution (R $\sim$~1000) grating instrument for the Palomar 200-inch (P200) Hale telescope \citep{Oke_1982}. The first spectrum for AT 2023tkw was taken with the DBSP on October 5, 2023. It has a blue camera observing in the range of 3400-5600~\AA~and a red camera for 5650-10500~\AA~region. The spectrum was wavelength-calibrated using arc lamps and flux-calibrated using the standard star Feige~34. The spectra were processed with the \texttt{python} package \texttt{dbsp\_drp} \citep{MandigoStoba_2022} implemented within the \texttt{pypeit} framework. 

\subsubsection{HCT/HFOSC}

The Himalayan Chandra Telescope (HCT), located at the Indian Astronomical Observatory (IAO), Hanle, is equipped with a low-resolution spectroscopic instrument, the Himalayan Faint Object Spectrograph Camera (HFOSC)\footnote{\href{https://www.iiap.res.in/centers/iao/facilities/hct/hfosc/}{https://www.iiap.res.in/centers/iao/facilities/hct/hfosc/}}. It was used to obtain the spectra of the transient AT~2023tkw with grism 7 in the wavelength range 3500-7800 \AA. FeAr lamp was used for wavelength calibration. Flux calibration was carried out against Feige~110 for observations on October 18 and 26, whereas Feige~34 was used for the observation on November 19. The two spectra taken on October 26 were similar and hence combined to increase the signal-to-noise ratio (SNR).

\subsubsection{Keck/LRIS}
The Low-Resolution Imaging Spectrometer (LRIS) is an optical wavelength imaging and spectroscopy instrument operating at the 10-m telescope, Keck I \citep{Oke_1995}. It observed the object just after the primary peak. The blue and red dispersion regions are 3000-5650~\AA~and 5650-10250~\AA, respectively. The spectrum was wavelength-calibrated using standard arc lamps and flux-calibrated using the standard star G191-B2B. The spectra were processed with the \texttt{idl} package \citep{Perley_2019} and were flux-calibrated using standard star observations.

\subsubsection{Keck/NIRES}
Near-Infrared Echellette Spectrometer (NIRES) is a prism cross-dispersed near-infrared spectrograph mounted on the 10m Keck II telescope \citep{Wilson_2004}. It observed the object in the wavelength region corresponding to the $JHK$ bands.
The spectra were reduced, extracted and wavelength calibrated using sky emission lines with the \texttt{idl} tool \texttt{spextool} \citep{Cushing_2004} and were telluric corrected and flux-calibrated using \texttt{xtellcor} \citep{Vacca_2003}.

\subsection{Extinction Correction}
\label{subsec:extinction}
The extinction towards AT~2023tkw was calculated from a 1$\arcmin$ region around the nova in the M160 resolution dust map of M31 \citep{Draine_2014} following a correction factor outlined in \cite{basu_2024a}. The obtained $A_V=0.621$, and $R_V=3.1$ were used to deredden all the optical-IR spectra. We used CCM89 extinction law form \cite{Cardelli_1989} in the \texttt{extinction} package \citep{Barbary_2021} to correct for the extinction. 


\section{Light and color curves}
\label{sec:lc_evol}
\begin{figure*}
    \centering
    \includegraphics[width=\linewidth]{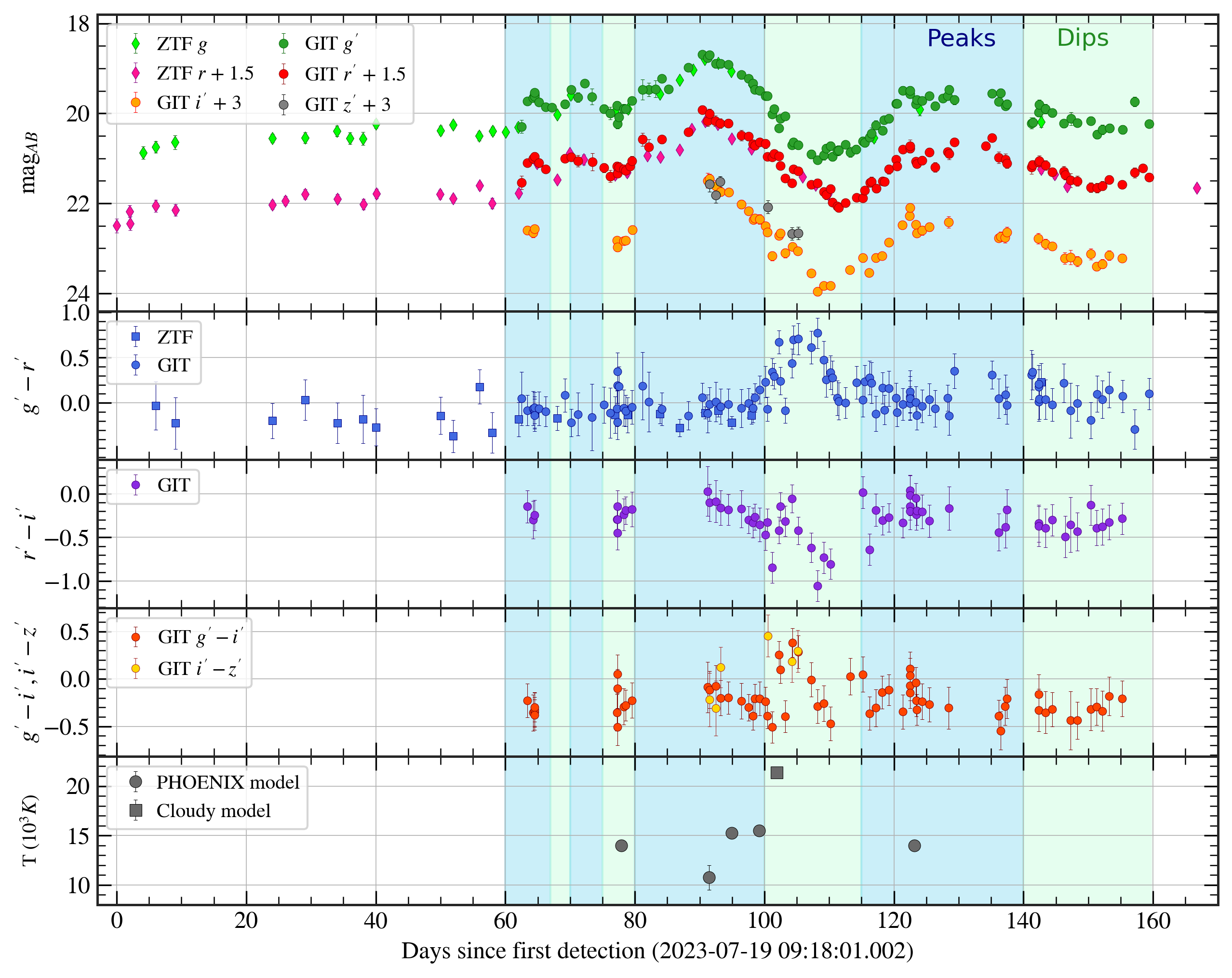}
    \caption{Optical light curve and extinction corrected color curves. The temperature evolution. obtained from spectral modeling, is shown in the last panel. The data associated with this figure will be available online.}
    \label{fig:optical_lc}
\end{figure*}

\subsection{Features and Classification}

The ZTF light curve (see Figure~\ref{fig:optical_lc}) provides information about pre-discovery evolution from 19 July 2023. It shows modulations and rises slowly by 1 mag in 60 days. It then showed a steep rise of $\sim$ 1~mag in 5~days, followed by a drop of more than 0.5 mag in the next five days. Immediately, a second sequence of \textit{peak and dip} was observed in the next 10~days. It rose again from $\sim$ 20.25 mags and attained the primary peak at $m_{\mathrm{peak}}= 18.5$~mags in 12~days, 90~days after its first detection. The extinction corrected peak absolute magnitudes reached by the object are $-6.52~(g^{\prime})$, $-6.58~(r^{\prime})$, and $-6.40~(i^{\prime})$. These are typical of novae in M31 but narrowly towards the fainter end \citep{Shafter_2009}. We should point out that M31 was in solar conjunction before mid-July, causing uncertainty in determining the exact time of the eruption. However, there is no doubt about the slow rise of the object from its light curve. M31 novae with jitters have been seen to brighten up slowly, than those with a smooth rise \citep{Cao_2012}. The nova then headed towards a dip of 2.5~mags in the next 20 days. After that, it started rising again, up to almost 2 mags and sustained a jittery plateau for around 15 days. It then showed another dip, this time at a slower rate, and faded to 20.5 mags. Our light curve coverage ends with a last rise at $\sim$ day 155-160 before finally fading away. The light curve resembles a J-class nova from \cite{strope_2010}. It is quite similar to the historical nova RR Pic \citep{Lunt_1926} with a slow and ragged rise, a similar drop followed by multiple rises. Another contemporary example of a slow nova with a rare gradual rise and multiple peaks is Gaia22alz (or Nova Velorum 2022; \citealt{Aydi_2023}). However, before drawing any comparison, we should note that the slow rise, in the case of AT 2023tkw, is restricted only to the final 2-3 mags before the peak. Prior information about the evolution from quiescent is unavailable.

A linear fit to the decline of $g^{\prime},\ r^{\prime},$ and $i^{\prime}$ light curves from the primary maxima to the $m_{\mathrm{peak}}+2$ value yields 16.0, 20.2, and 15.9 days respectively. Such a short $t_2$ time might suggest the object to be a \textit{fast} nova \citep{Warner_2003}. But novae with J class light curves can cross the $m_{\mathrm{peak}}+2$ threshold more than once. For such cases, \cite{strope_2010} used the last time when the light curve crossed the $m_{\mathrm{peak}}+2$ to determine the $t_2$ time. AT~2023tkw reaches close to but doesn't cross the $m_{\mathrm{peak}}+2$ value in $g^{\prime}, r^{\prime}$ bands for the second time during its decline at day $\sim$ 60 from the primary peak, whereas the $i^{\prime}$ band crosses it at day 61 again, although not for the last time. Due to scarce coverage beyond day 160 since first detection (see Figure~\ref{fig:optical_lc}), we could only put constraints on the $t_2$ time being $t_2>75$~days, confirming the \textit{slow} nature of the nova. 

The $g^{\prime}-r^{\prime}$ color becomes redder during the major dip at day 100-115 as shown in Figure~\ref{fig:optical_lc}. The color also seems to become redder during the other minor dips in Figure~\ref{fig:optical_lc}, but the change is within the error bars. A likely cause for this is enhanced H$\alpha$ emission, which tends to dominate as the continuum flux decreases. The $r^{\prime}-i^{\prime}$ becomes blue during the major dip because of the same reason as above. The $g^{\prime}-i^{\prime}$ color, on the other hand, is devoid of the strong H$\alpha$ flux. During the decline around day 100-115, it becomes redder before becoming bluer at the light curve minima at day 110.


\section{Spectroscopy}
\label{sec:spec_evol}
\subsection{Spectral Evolution}

\begin{figure*}
    \centering
    \includegraphics[width=0.9\linewidth]{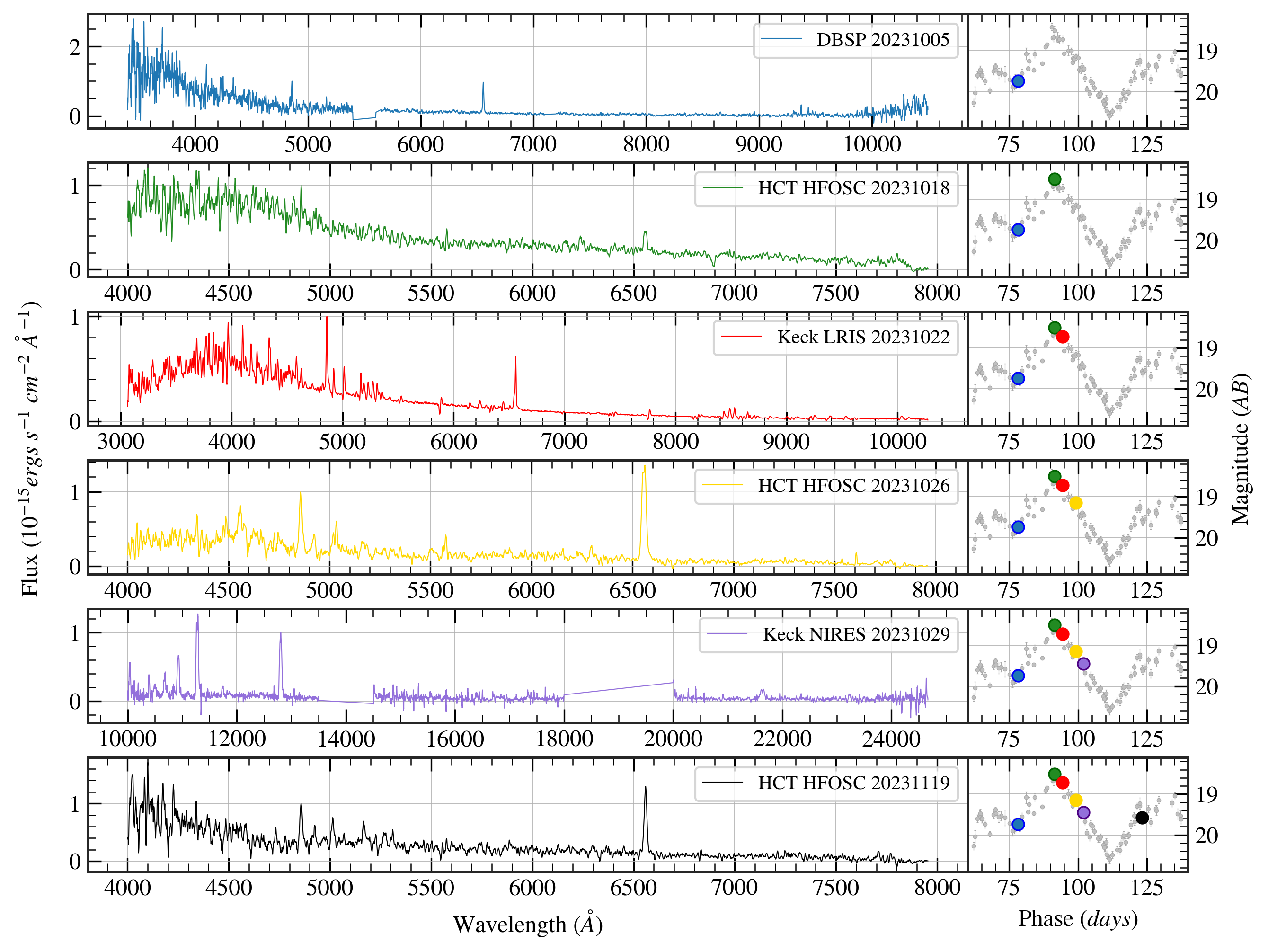}
    \includegraphics[width=0.85\linewidth]{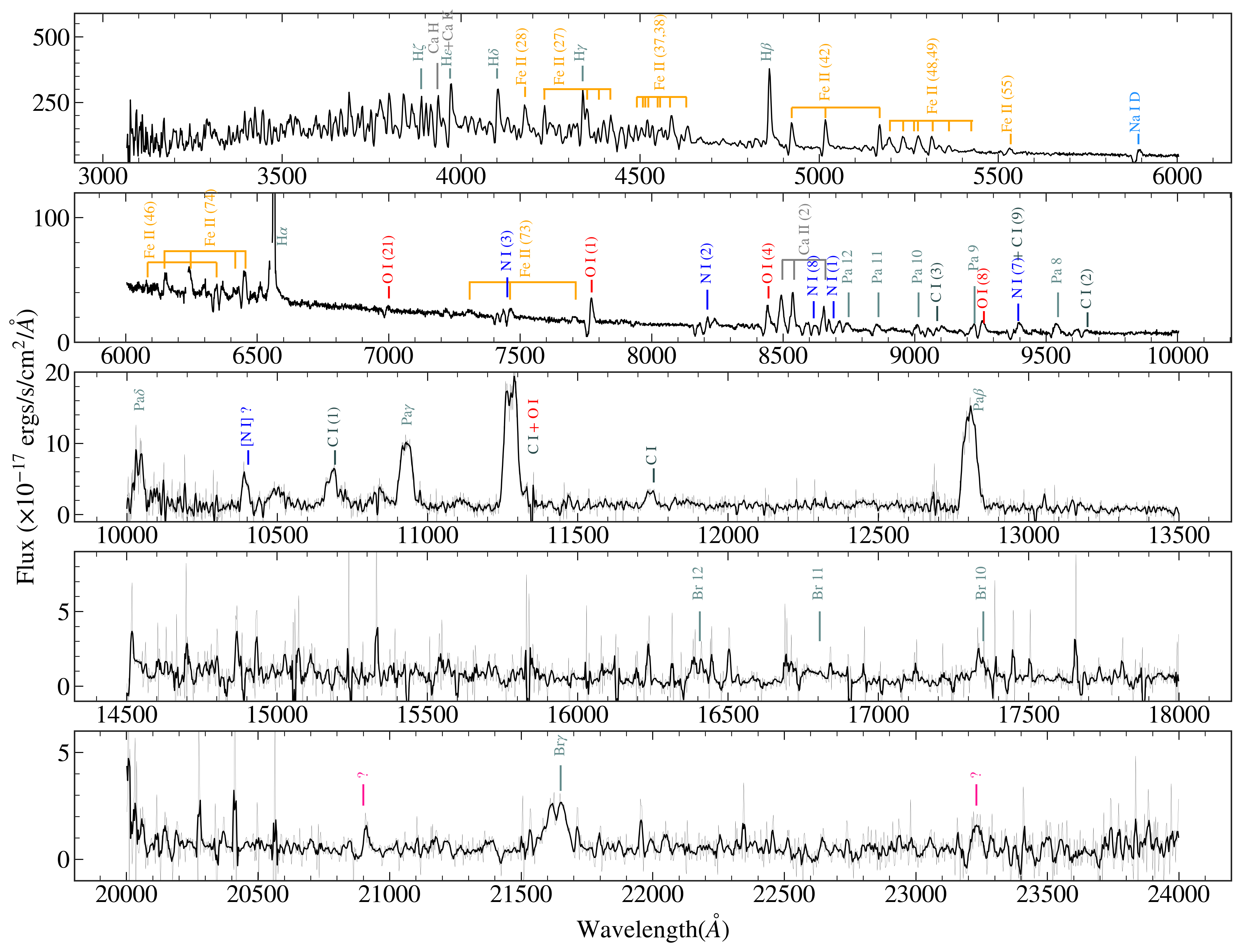}
    \caption{\textit{Top:} Spectroscopic evolution of AT~2023tkw. The spectral epochs are overplotted on the light curve in the right-hand panel. \textit{Bottom:} Line identification in the Keck LRIS and NIRES spectra.}
    \label{fig:spec_plot}
\end{figure*}

The sequence of spectra shown in Figure~\ref{fig:spec_plot} was observed at different phases of the light curve, i.e., at dips, primary maxima, during the decline, and during the second maxima using different instruments. H Balmer lines are visible in all the epochs in the optical spectra. The line fluxes were initially high during the rise, and the continuum was blue in the first epoch, as seen in the top panel of Figure~\ref{fig:spec_plot}. The pre-maxima widths of the Balmer lines were 350~$\mathrm{km~s^{-1}}$, indicating a slowly expanding ejecta. At the peak of the light curve, the continuum dominated the optical bands. The emission lines weakened in intensity and broadened to more than 600~$\mathrm{km~s^{-1}}$. However, this could be overestimated as it is at the edge of HFOSC's resolution. During its decline to the first dip, the H line widths in the LRIS spectrum decreased to FWHM$\approx$400-500~$\mathrm{km~s^{-1}}$. Almost all the emission lines in the LRIS spectrum are accompanied by superimposed P~Cyg absorption features at roughly $-700$~$\mathrm{km~s^{-1}}$ from the emission peaks (3rd panel in Figure~\ref{fig:optical_lc}). The emission lines progressed to increase in strength and in width, up to 1000~$\mathrm{km~s^{-1}}$ measured in the HFOSC and NIRES spectra (4th and 5th panel in Figure~\ref{fig:spec_plot}). At the same time, the peak of the SED shifted to shorter wavelengths, manifesting flatter spectra in the optical and IR bands, beyond ~5000~\AA. The \ion{Fe}{2} (42,48,49) lines can be seen in the principal spectra. The second peak was characterized by similar features as in the first peak, i.e., dominated by a blue continuum, weakened emission line strength, and velocities reducing to 500~$\mathrm{km~s^{-1}}$. The observed spectral evolution aligns with the typical pattern for novae displaying multiple peaks.

The Keck LRIS spectrum shows numerous H Balmer, H Paschen, and \ion{Fe}{2} multiplets, along with Ca, \ion{O}{1}, \ion{N}{1} and \ion{C}{1} lines identified in Figure~\ref{fig:spec_plot}. The P Cyg absorption component of the Balmer, \ion{Fe}{2}, and \ion{O}{1} lines are suggestive of a dense shell moving outwards at a velocity of 600-800~km s$^{-1}$. The IR spectrum displays prominent Paschen and Brackett lines, as well as \ion{C}{1} and \ion{O}{1} lines marked in Figure~\ref{fig:spec_plot}. The line lists given in \cite{Williams_2012} and \cite{banerjee_2012} were used to identify the emission features in the optical and IR spectra, respectively.

The \ion{O}{1} $\lambda$11287~\AA~ and $\lambda$8446~\AA~are seen to be enhanced by fluorescence due to Lyman $\beta$ pumping, which is commonly seen in novae \citep{Kastner_1995}.


\subsection{Spectral modelling}
We used the \texttt{PHOENIX} (\citealt{Husser_2013} and references therein) and \texttt{Cloudy} (\citealt{cloudy17} and references therein) codes to model the optically thick and optically thin phases of the nova, respectively.
\begin{figure}
    \centering
    \includegraphics[width=\linewidth]{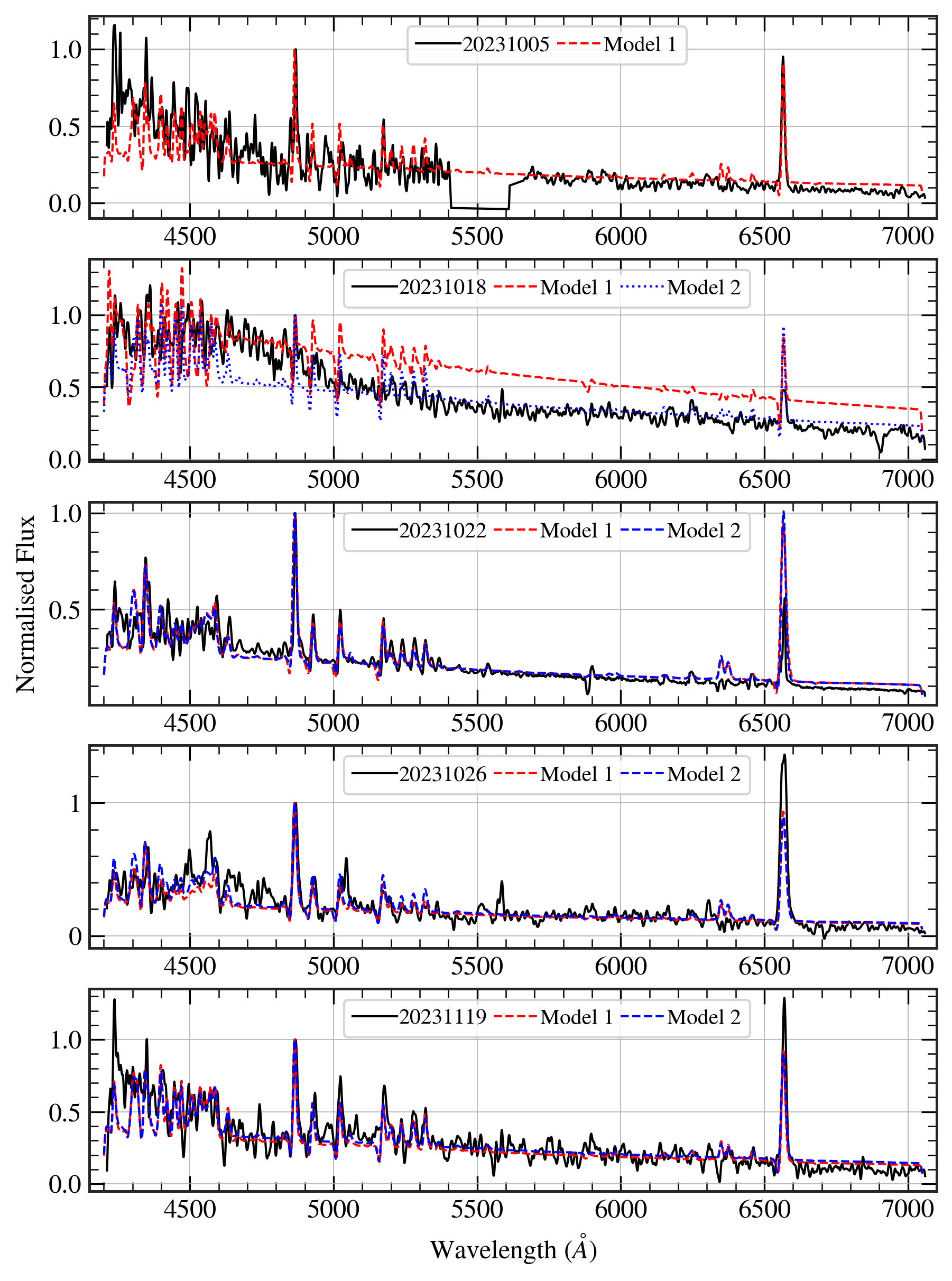}
    \includegraphics[width=\linewidth]{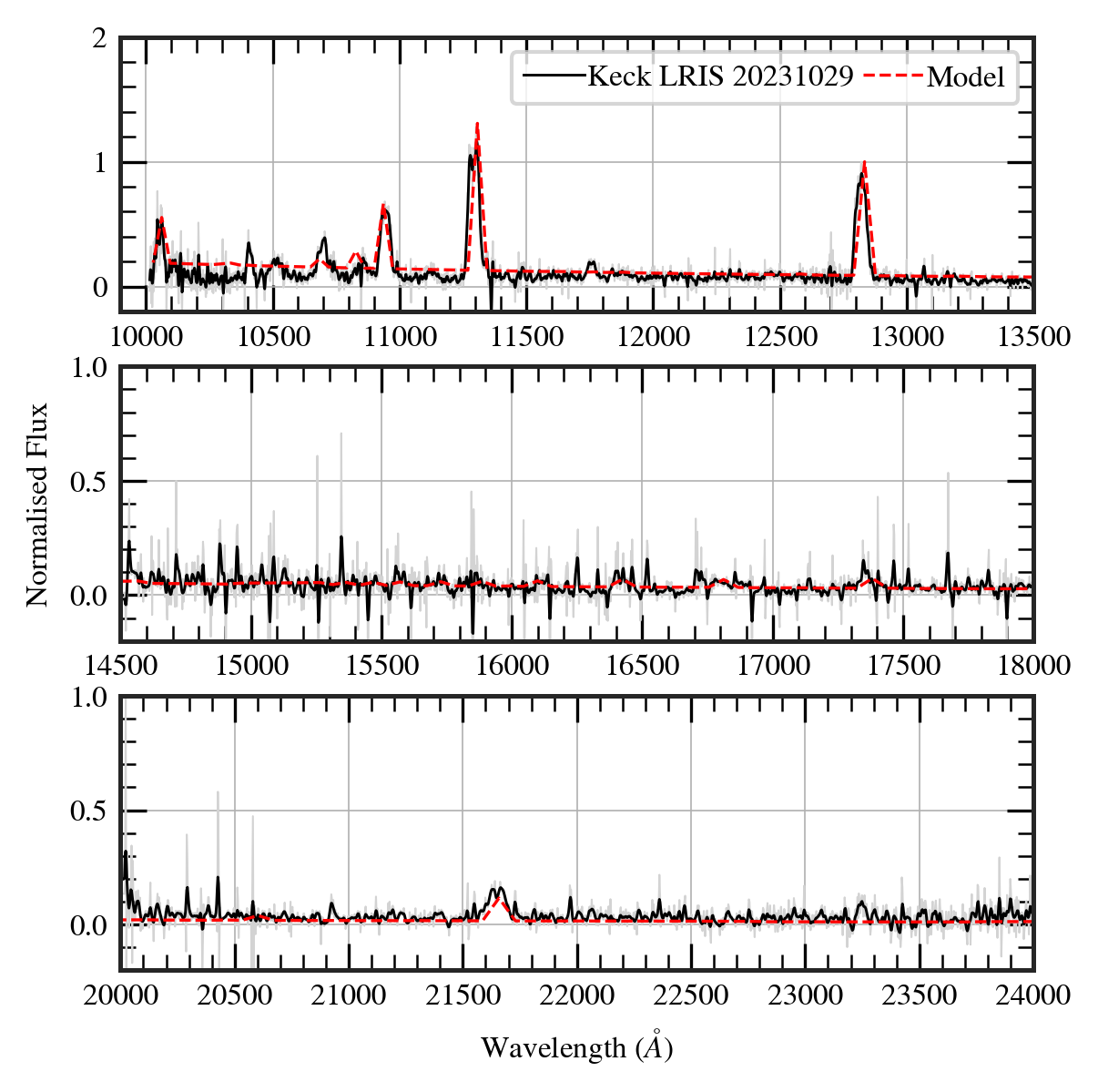}
    \caption{\textit{Top:} \texttt{PHOENIX} models overplotted on the optical spectra. \textit{Bottom:} \texttt{Cloudy} model of the IR spectrum overplotted on the observed spectrum.}
    \label{fig:spectra_model}
\end{figure}

\begin{table}[]
    \centering
    \caption{\texttt{PHOENIX} model parameters of optical spectra and \texttt{Cloudy} model parameters of the IR spectrum observed on 2023-10-29.}
    \label{tab:phoenix_model}

    \begin{tabular}{ccccc}
    \multicolumn{5}{c}{\texttt{PHOENIX} model parameters} \\
    \toprule
       Date & Phase & $T$ & $v_0$ & Metallicity \\
       (UT)   &     & ($10^3$ K) & ($\mathrm{km~s^{-1}}$) &  \\ 
    \midrule 
        2023-10-05 & Pre-maxima & 14.0      & 1000      & 0.0 \\
        2023-10-18 & Peak       & 9.5-12.0  & 1000-1500 & 0.0 \\
        2023-10-22 & Decline    & 15.0-15.5 & 1000-1500 & 0.0 \\
        2023-10-26 & Decline    & 15.5      & 1500      & 0.0-0.2 \\
        2023-11-19 & Peak       & 13.5-14.5 & 1000-1500 & 0.0-0.2 \\
    \bottomrule
    \end{tabular}

    \bigskip

    \label{tab:cloudy_model}
    \begin{tabular}{lc}
    \multicolumn{2}{c}{\texttt{CLOUDY} model parameters} \\
    \toprule
    Parameter  &  Value \\
    \midrule
    Ionising Source & Blackbody \\
    $\mathrm{T_{BB}}$   & 21380 K \\
    Luminosity & $10^{38}$ $\mathrm{ergs~s^{-1}}$ \\
    Hydrogen density &  $7.94 \times 10^8$ $\mathrm{cm^{-3}}$\\
    Density Parameter N & -3 \\
    Inner radius & $5.25 \times 10^{14}$ $\mathrm{cm}$ \\
    Outer Radius & $10.96 \times 10^{14}$ $\mathrm{cm}$ \\
    Filling factor & 0.1 \\
    $\mathrm{C/C_{\odot}}^a$  &  37.17 \\
    Ejected Mass   & $8.92\times 10^{-5}~\mathrm{M_{\odot}}$   \\
    \bottomrule
    \end{tabular}
    \\
    $^a$ The solar abundance of C relative H was taken to be $2.69\times 10^{-4}$ from \cite{GASS10}.

\end{table}

\subsubsection{Optically thick phase}
\label{phoenix}
\texttt{PHOENIX} code was developed to model stellar photospheres, including optically thick expanding atmospheres of novae and supernovae \citep{Hauschildt_1995}. The models are generated under NLTE conditions by solving spherically symmetric special relativistic radiative transfer equations \citep{Hauschildt_1997}. These include thousands of transitions, including Fe, C, O, and N lines, and span a wide range of temperatures and velocities to cover the evolution of novae spectra \citep{Hauschildt_1997}. The models are characterized by the model temperature ($T$), the luminosity ($L$) (connected to radius of photosphere $R$ by Stefan-Boltzmann's law), the density parameter $N$ in $\rho \propto r^{-N}$, the mass loss rate ($\dot M$), the velocity  ($v_0$) of the expanding photosphere at $R$, and the metallicity $z$ \citep{Schwarz_1997}. The luminosity of the central star was fixed at 50000 $L_{\odot}$, typical of outbursting novae (see Figure~1 in \citealt{Kato_2024}). Previous studies have shown that $N=3$ fits nova spectra well \citep{Schwarz_1997, Schwarz_1998, Schwarz_2001}. With an assumption of constant $\dot M$, the velocity field becomes linear with radius. Models were generated by varying $T$, $v_0$, and $z$. A grid search was performed for $T$ in the full range of [6000, 30000]~K, $v_0$ in [1000, 1500]~$\mathrm{km~s^{-1}}$ and $z$ in [0, 0.4]. The velocity was constrained by the width of emission lines in the observed spectra.

The models closely resembling the optically thick spectra obtained during the rise up to the initial decline and during the secondary peak are displayed in Figure~\ref{fig:spectra_model}, and the parameters are given in Table~\ref{tab:phoenix_model}. The models can reproduce the continuum and the emission features in all the optical spectra observed at different epochs except the one taken at the primary peak around day 90 from the first detection. The P Cyg features are also well reproduced in the models. A low temperature at the optical maxima aligns with the picture, indicating that the photosphere is at its maximum expansion during the optical peak. Further evolution of the temperature (see the bottom panel in Figure~\ref{fig:optical_lc}) signifies ejecta becoming optically thin, revealing the shrinking photosphere exposing the inner hotter layers of the system. The H$\alpha$ line in the Keck/LRIS spectrum has been ignored while comparing with the models, as the line peak is saturated. At the phase on October 26 ($\sim$~8~days from peak), we can see that the observed H$\alpha$ flux exceeds the atmospheric model prediction, while other lines fit relatively better. This excess H$\alpha$ emission could be coming from the part of ejecta, which has decoupled from the photosphere and evolved to an optically thin state. It could also be coming from the ionized circumbinary material.

We also tried to model the IR spectrum with \texttt{PHOENIX}, but the continuum could not be matched for this case. Another issue was reproducing the fluorescence line from \ion{O}{1} $\lambda$11287~\AA. \texttt{PHOENIX} underestimated the line strength across the temperature, velocity, and metallicities grid. 

The spectrum taken at the secondary peak is again well fit, both in terms of continuum and emission lines by \texttt{PHOENIX}. H$\alpha$ is observed to be enhanced at the second maxima. This could be the emission from the already photoionized ejecta from the previous mass ejection episode. The drop in temperature is also consistent with the photosphere expanding again during the second peak. 

\subsubsection{Optically thin phase}
Motivated by the booming H emission lines and expansion of the ejecta during the decline phase, we decided to model the NIRES spectrum taken 11 days after the peak using the photoionization code \texttt{Cloudy} (\texttt{v22.0}; \citealt{cloudy17, cloudy23}). \texttt{Cloudy} is capable of generating synthetic spectra of a gas cloud irradiated by a hot source. The output spectrum is calculated from ionization, temperature, and density distribution in the cloud. A 1D model was generated using a single cloud component in the spherically expanding ejecta. The free parameters in a model were the black body temperature of the source, and the hydrogen density and abundance of the ejcta. The ionizing source was found to be a black body radiating at a temperature of 21380~K with a luminosity of $10^{38}$ $\mathrm{ergs~s^{-1}}$. The density parameter $N=-3$ was chosen to satisfy a ballistic expansion, like in \texttt{PHOENIX}. The inner and outer radii of the shell were calculated from the emission line velocities, assuming the first detection as the eruption date. All the parameters were chosen following the methods in \citealt{Pav19} (and references therein) and are mentioned in Table~\ref{tab:cloudy_model}. The above parameters with an H density of $\sim 8\times 10^8$~$\mathrm{cm^{-3}}$ could match the Paschen and Brackett lines in the NIR spectrum shown in Figure~\ref{fig:spectra_model}. A high carbon abundance of $\mathrm{C/C_{\odot}}=37$ was required to model the observed C lines. An overabundance of C and other elements has been found in novae ejecta due to ablation of material from the WD surface during the TNR \citep{Gehrz_1998}. The ejecta mass was estimated to be of the order of $10^{-4}~\mathrm{M_{\odot}}$. The temperature prediction from \texttt{Cloudy} modeling is consistent with a hot ionizing source, achieved when the ejecta has become optically thin and geometrically thick.


\section{The CV system}
\label{sec:progenitor}

\subsection{Archival non-detections}
We combined images of the nova field from the GIT archives to search for the nova system at quiescence. We selected ten images from both $g^\prime$ and $r^\prime$ bands, each of 240~s exposure and captured under the best seeing conditions (FWHM~$\approx~2-3\arcsec$), between July and September 2022. Two sets of co-added images were generated by performing a median and sum combination in \texttt{SWarp} \citep{swarp}. No source was detected in either set at a 3$\sigma$ limiting AB magnitude of 21.3 and  21.2 in the $g^\prime$ and $r^\prime$ bands, respectively.

The ZTF forced photometry service was also used to obtain archival upper limits for the source between 30 May 2018 and 16 July 2023, covering $\sim850$ and $\sim1150$ epochs in the ZTF-$g$ and ZTF-$r$ bands respectively. The median and standard deviation of upper limit achieved was $20.9~\pm~0.6$ ($g$) and $20.7~\pm~0.6$ ($r$), with the deepest limiting magnitude of $21.97$ ($g$) and $21.72$ ($r$).

No sources were detected in archival AstroSat Ultraviolet Imaging Telescope (UVIT; \citealt{KPS_2014, Tandon_2020}) images observed on 8 November 2017, up to a limiting magnitude of 22.83 in F148W, 22.07 in F172M, 22.36 in N219M, and 21.87 in N279N filters. 

The Swift XRT 2SXPS upper limit server\footnote{\url{https://www.swift.ac.uk/2SXPS/ulserv.php}} \citep{Evans_2020} was used to obtain a limit of $<~3.1~\times~10^{-4}$~counts~s$^{-1}$ in the $0.3 - 10.0$~keV range, between 1 September 2006 and 26 July 2018.

We looked for the object again in the GIT images once M31 was visible in the later half of 2024. The object was not detected up to a limit of 18.9~mags in the GIT $r^{\prime}$ images taken on 10 August 2024.

\subsection{HST photometry}
\label{subsec:hst_phot}
\begin{figure}
    \centering
    \includegraphics[width=\linewidth]{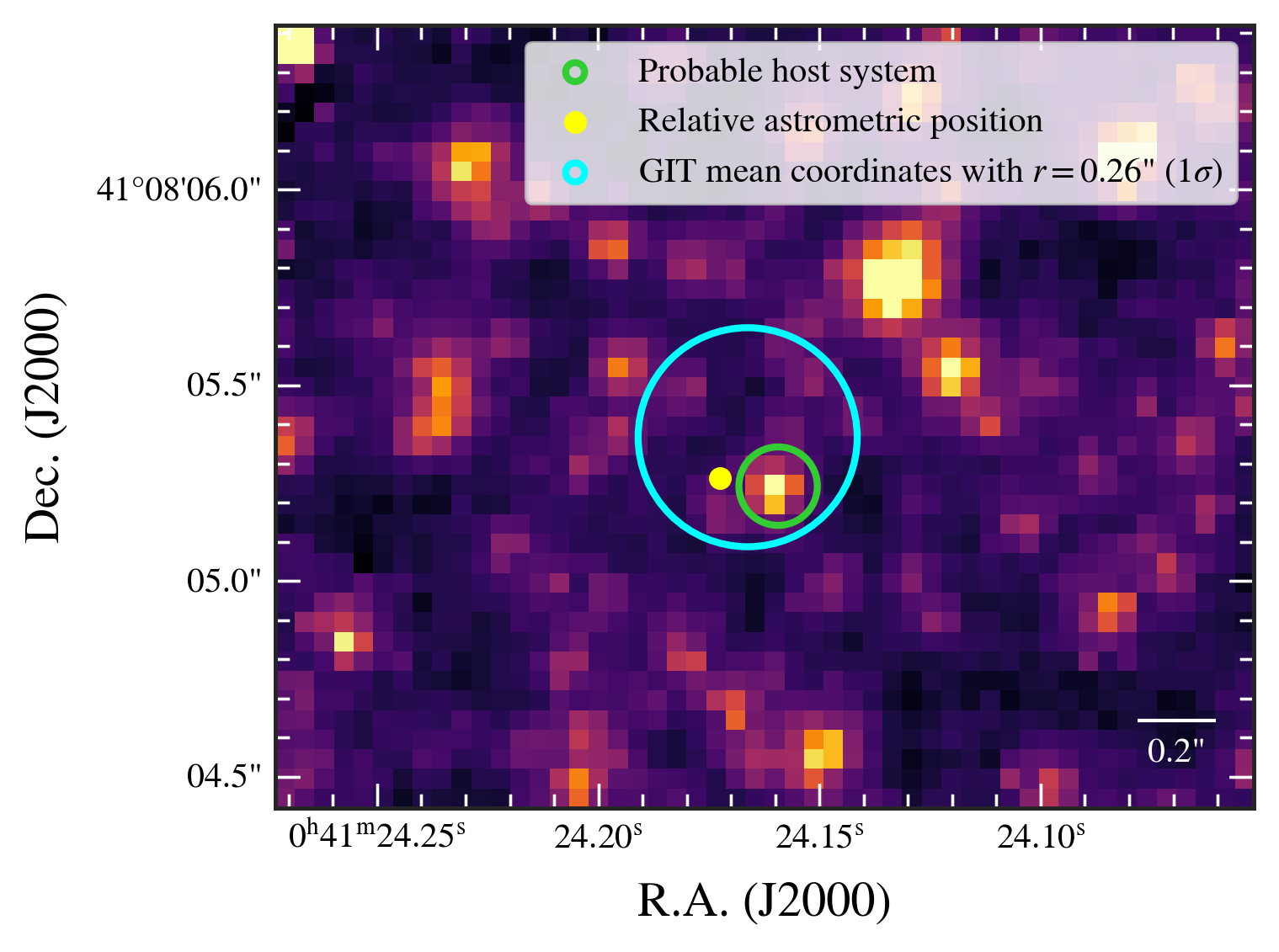}
    \includegraphics[width=\linewidth]{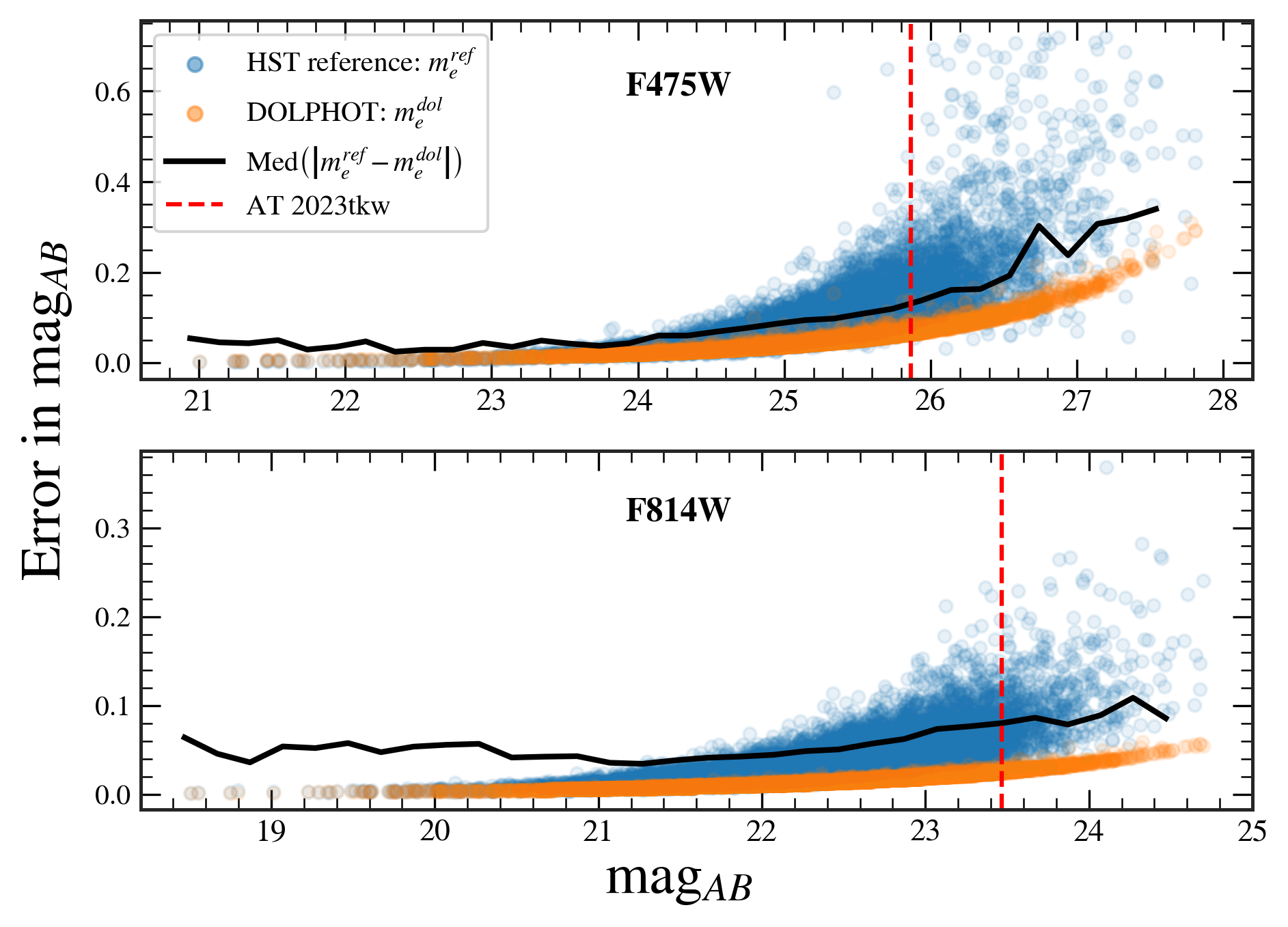}
    \caption{\textit{Top:} HST ACS/WFC image of the likely host system, marked in the green circle in the F814W filter. The coordinate of AT2023tkw obtained by relative alignment with the HST image is marked by a yellow dot. The GIT localization and its uncertainty is marked by a cyan circle. }\textit{Bottom:} Photometric errors as a function of magnitude, obtained using the provided HST reference catalog magnitude error ($m^{\text{ref}}_{e}$) and the \texttt{DOLPHOT} catalog errors ($m^{\text{dol}}_{e}$).
    \label{fig:dolphot}
    \label{fig:progenitor}
\end{figure}

The Hubble Space Telescope (HST) observed the field on 27 December 2022 during the HST cycle 29 under the proposal GO-16796 (PI: Benjamin Williams). The images were taken using Hubble's Advanced Camera for Surveys' ACS; \citealt{Ryon_2023, Hathi_2024}) Wide Field Channel (WFC) in the F475W and F814W filters, with an effective exposure time of $\sim$1000~s in each. 

Accurate astrometry and identification of the source in the dense stellar field (Figure~\ref{fig:progenitor}) was performed by resampling images from the GIT using \texttt{SWarp} \citep{swarp} to the ACS/WFC resolution of 0.05~arcsec~pixel$^{-1}$ and then registering them to the HST images using \texttt{Spalipy} \citep{spalipy}. A source in the HST/ACS images, consistent within the GIT 1$\sigma$ error circle ($r=0.26$\arcsec) of AT~2023tkw, was found at the sky coordinates ($\alpha$, $\delta$) = (00:41:24.1638, +41:08:05.243) as shown in the top panel of Figure~\ref{fig:progenitor}. The source density near the object in the HST/ACS image was found to be 0.56~source/arcsec$^2$. Thus the probability of a field star being present within the GIT error circle is rather low at $11.89\% $. However, since we do detect a source in the HST/ACS images within the GIT error circle, it is quite likely that the source is the host system of AT~2023tkw. However, we should note that there is a non-zero chance that this source is not the host and the system at quiescence is fainter than the detection limit.

\texttt{DOLPHOT} \citep{dolphot} was used to perform photometry of the images. The ACS/WFC pixel area maps and corresponding PSFs for the filters were used for photometric calibration, following the standard steps described in \cite{hstphot}. We filtered out \texttt{DOLPHOT} sources that had a $\texttt{crowding}~>~0.2$ to remove poorly measured stars. The remaining sources were then cross-matched with sources in the HST catalog\footnote{\href{http://dx.doi.org/10.17909/b4vg-pa17}{http://dx.doi.org/10.17909/b4vg-pa17}} for each image, and the median photometric errors were calculated as a function of magnitude (see bottom panel in Figure \ref{fig:dolphot}). 
We obtain the AB magnitudes of the (potential) nova host system to be $25.75~\pm~0.13$ in the F475W filter and $23.89~\pm~0.08$ in the F814W filter. Correcting for the extinction calculated in \S\ref{subsec:extinction} and assuming a distance of 778~kpc \citep{Stanek_1998}, the absolute magnitudes are $0.66~\pm~0.13$ and $-1.35~\pm~0.08$ in the F475W and F814W filters, respectively.


\section{Discussion}

\subsection{Binary parameters}
\label{subsec:binary_param}
From the ZTF light curve on days 0-60, we tried to determine if any periodicity was associated with it. However, the periodogram signal was weak, and the results were inconclusive. 

The extinction-corrected quiescent absolute magnitudes of the potential host system detected in the HST/ACS images are $0.66~\pm~0.13$ (F475W) and $-1.35~\pm~0.08$ (F814W). These are consistent with giant stars (sub-giants/main sequences being fainter by $>$2.5 mags). Converting to $B,I$ mags using the suitable transformations \citep{Sirianni_2005, Harris_2018}, we obtained the color $B-I = 3.07$. This resembles cool K-type evolved companions whose $B-I$ values are around $\sim$2.3 (K0~III), $\sim$2.6 (K2~III), and $\sim$3.6 (K5 III). We should note that the F475W filter can likely have contributions from the accretion disk; thus, the secondary could be an even cooler star.

The estimated ejecta mass from the \texttt{ cloudy} modeling of the NIR spectrum yielded a value of $\sim 9\times 10^{-5} \mathrm{M_{\odot}}$. This is on the higher end of the ejecta mass, but this is not unusual for a classical nova. It is close to the ones observed for slow novae with a low mass CO WD (eg. V723 Cas: \citealt{Evans_2003}). 

Upon comparing the observed properties of the outburst, that is, $\mathrm{M_{ej}} \approx 10^{-4}~\mathrm{M_{\odot}}$ and $\mathrm{v_{exp}} \approx 700~\mathrm{km~s^{-1}}$, with models from \cite{Yaron_2005}, the closest WD parameters that match are around $\mathrm{M_{WD}} = 0.65~\mathrm{M_{\odot}}$, $\mathrm{\dot M} = 5 \times 10^{-13}~\mathrm{M_\odot yr^{-1}}$, and $\mathrm{T_{WD} = 30~MK}$. Such systems evolve very slowly, which is consistent with the observations. Their recurrence timescale is of the order of hundreds of millions of years.

The low accretion rate and long evolution timescale might seem unusual for novae systems hosting evolved secondaries. However, there have been instances of CNe, such as V841 Oph, CP Cru, GK Per, and V723 Cas, with evolved secondaries \citep{Warner_2008}. In fact, V723 Cas shows a slow rise, and the decline comprises multiple peaks \citep{kiyota_2004}, similar to AT~2023tkw. Some more examples can be found in \citealt{Mroz_2015} (V2674 Oph, V5589 Sgr) and \citealt{Schaefer_2021} (FM Cir, V392 Per, V1534 Sco, V5583 Sgr). HV Cet \citep{Beardmore_2012}, V1708 Sco \citep{Sokolovsky_2020}, and V3732 Oph \citep{Mroz_2021} are some recent CN systems hosting evolved secondaries. \cite{Mroz_2015} also mentioned the presence of two peaks in the orbital-period distribution of CN, corresponding to main-sequence and evolved secondaries.

\subsection{Cause of multiple peaks in the light curve}
AT 2023tkw light curve exhibits two brightenings during its rise to the primary peak. A broader secondary peak is seen 30-45 days after the primary maximum (Figure~\ref{fig:optical_lc}). Several novae have shown re-brightenings in their light curves (for example, V723 Cas: \citealt{kiyota_2004}; V2540 Oph: \citealt{Ak_2005}; V1548 Aql, V4745 Sgr: \citealt{Csak_2005}; V1186 Sco: \citealt{Schwarz_2007}; V458 Vul: \citealt{Poggiani_2008}; V5558 Sgr: \citealt{Poggiani_2008a}; V1405 Cas, V606 Vul, FM Cir \citealt{Aydi_2024}). 

\cite{Cao_2012} and \cite{Clark_2024} reported many M31 classical novae from their respective surveys with re-brightening features in their light curves before and/or after the primary peak. All of these are classical novae and have a slow rise time. The distribution between smooth- and jittery- class novae in M31 was broadly consistent with that in the Milky Way. Below, we discuss some causes of the observed rebrightenings.

\subsubsection{A case of dust dip?}
The dip in the light curve at day 110 may resemble a ``dust dip'' as seen in Figure~\ref{fig:optical_lc}. But such a feature is usually seen 50-100 days after the peak during the $t_3$ transition timescale \citep{strope_2010}. On the contrary, the dip is noticed at 10 days from the peak here, earlier than the $t_2$ time. Further, dust dips are generally observed to be deep, sometimes as deep as 10~mag (D-class light curves). \cite{strope_2010} also argues that the shallow dips seen in some D-class light curves (such as  OS And, V476 Cyg, and NQ Vul) may not be caused by dust. AT~2023tkw does show a comparatively shallow dip of 2.5~mag. \cite{Shafter_2011} reported from the Spitzer survey of M31 about the IR excess during dust formation and the correlation of dust condensation timescale with the speed class of novae. Theoretical models of \cite{Williams_2013} also backed up the dependency of dust formation on the speed class of a nova. For a slow nova, like AT 2023tkw, the dip in magnitude is unlikely to be caused by dust. Dust dips are usually distinct features in light curves and can be seen as stark deviations from the decline, but the decline is smooth in this case. The lack of evidence of a red continuum in the color curves (Figure~\ref{fig:optical_lc}) and spectral evolution (Figure~\ref{fig:spec_plot}) also contradicts the dust formation scenario. Additionally, the light curves indicate the presence of another shallow dip around day 150. We, therefore, rule out dust formation as the cause of the dips in the light curve. 

\subsubsection{Transition from quasi-static to optically thick wind state?}
Another school of thought suggests that multiple peaks can be related to the binary separation and extent of the photosphere during the outburst. \cite{Kato_2009} found out from their simulations that novae eruptions can have two solutions: optically thick winds and quasi-static expansion. Massive WDs always evolve through optically thick winds and display a sharp peak. On the other hand, intermediate mass WDs (0.6-0.7 $\mathrm{M_{\odot}}$) can undergo either or both depending on the ignition mass. A quasi-static evolution would manifest as a flat-topped peak in the nova light curve, unlike that of a massive WD. The nova can transition from the static state to the optically thick wind state for the intermediate mass WDs. Such a shift is a violent process that releases energy, giving rise to jittery peaks seen in novae light curves. In their following work, \cite{kato_2011} further suggested that the close orbit of the secondary star could trigger this transition. Under such circumstances, the secondary is fully embedded inside the nova envelope during the eruption and can alter the internal structure of the envelope from a static expansion to an optically thick wind mass loss state. This phenomenon has been seen in the close binary V723 Cas. 

However, this situation is an unlikely scenario. As discussed in \S \ref{subsec:binary_param}, the secondary, being a giant star, is expected to be in a wide binary system. It decreases the possibility of the secondary altering the state of the envelope to give rise to multiple peaks.

\subsubsection{Photospheric oscillations?}

\cite{Tanaka_2011a} linked the appearance and disappearance of P Cyg profiles of V5558~Sgr during the maxima and minima in the light curve, respectively, to the expansion and contraction of the photosphere. The P Cyg velocity was also lower at the rebrightenings and higher during the decline/minima.. \cite{Aydi_2024} attributed the low-velocity P Cyg at the rebrightenings to the dense, slow outflow. In contrast, the high-velocity P Cyg away from the peaks is due to the intermediate (principal) component formed after a collision between the slow and fast outflows. 
\cite{Steinberg_2020} used 1D hydrodynamical models to show that multiple transitions between slow and fast outflows generated by internal shocks originating near or below the photosphere can result in multi-peaked nova light curves. The internal shocks temporarily increase the brightness of the system, seen as a peak in the light curve. As this shock front quickly reaches the photosphere, it stretches the photosphere and cools it before the photosphere contracts again, becoming hotter. Such alternate shocks can cause the photosphere to pulsate, changing the brightness and temperature of the source. 

Interestingly, the transient V612 Sct (ASASSN-17hx) shows a similar light curve with rebrightenings like that of AT 2023tkw \citep{Pavana_2020}. The rebrightenings were suggested to be caused by oscillations of the photosphere \citep{Mason_2020}. However, though this object had many novae-like properties, it was termed an imposter because of its massive ejecta $\sim 7-9\times10^{-4}~\mathrm{M_{\odot}}$, that has never been seen in novae. It was suggested that such slow ``novae'' originate from either very low mass WDs or some other transient phenomenon.

\subsubsection{Multiple ejections and shocks?}
\cite{Aydi_2024} suggested multiple ejection episodes as another possible explanation for multipeaked novae light curves. \cite{Csak_2005} mentioned that H burning on the surface of WDs is highly unstable, and expanding photospheres can lead to mass loss, i.e., secondary episodes of mass ejection. If dense shells of materials are expelled at intervals, it can lead to shocks due to the collision between the successive shells. These shocks superheat the plasma to millions of degrees, driving some charged particles to relativistic speeds. This process results in the generation of both thermal and non-thermal radiation. 

GeV $\gamma$-rays have been detected near the optical maxima from more than 20 novae \footnote{\href{https://asd.gsfc.nasa.gov/Koji.Mukai/novae/novae.html}{https://asd.gsfc.nasa.gov/Koji.Mukai/novae/novae.html}} (see \citealt{Chomiuk_2021} and references therein). Some of them, such as V1369~Cen and V5668~Sag \citep{Cheung_2016}, V357~Mus \citep{Gordon_2021}, V549~Vel \citep{Li_2020}, V1405~Cas \citep{Buson_2021} and V1723~Sco \citep{Cheung_2024, Cheung_2024b} display multi-peaked lightcurves. The first nova to show a remarkable correlation between $\gamma$-ray peaks and optical rebrightening was V5856~Sgr (ASASSN-16ma; \citealt{Li_2017}). The authors suggested that the majority of the optical emission might be reprocessed shock-heated plasma. With simultaneous optical and $\gamma$-ray light curves of V906~Car (ASASSN-18fv), \cite{Aydi_2020a} also found that high energy radiation is coupled with optical emission. The $\gamma$-ray peaks were detected $\sim$ 5.3 hours before the optical peak in V906 Car. Their spectra revealed the presence of multiple outflows and shock being generated in the ejecta when fast flow fronts collide with the slow flows.

However, it should be noted that there are exceptions, such as nova V549 Vel \citep{Li_2020}, where the $\gamma$-ray and optical light curves were not correlated, suggesting that the optical light may not be the reprocessed shocked emission necessarily.

\subsection{Increasing time gap between lightcurve peaks}

\begin{figure}
    \centering
    \includegraphics[width=\linewidth]{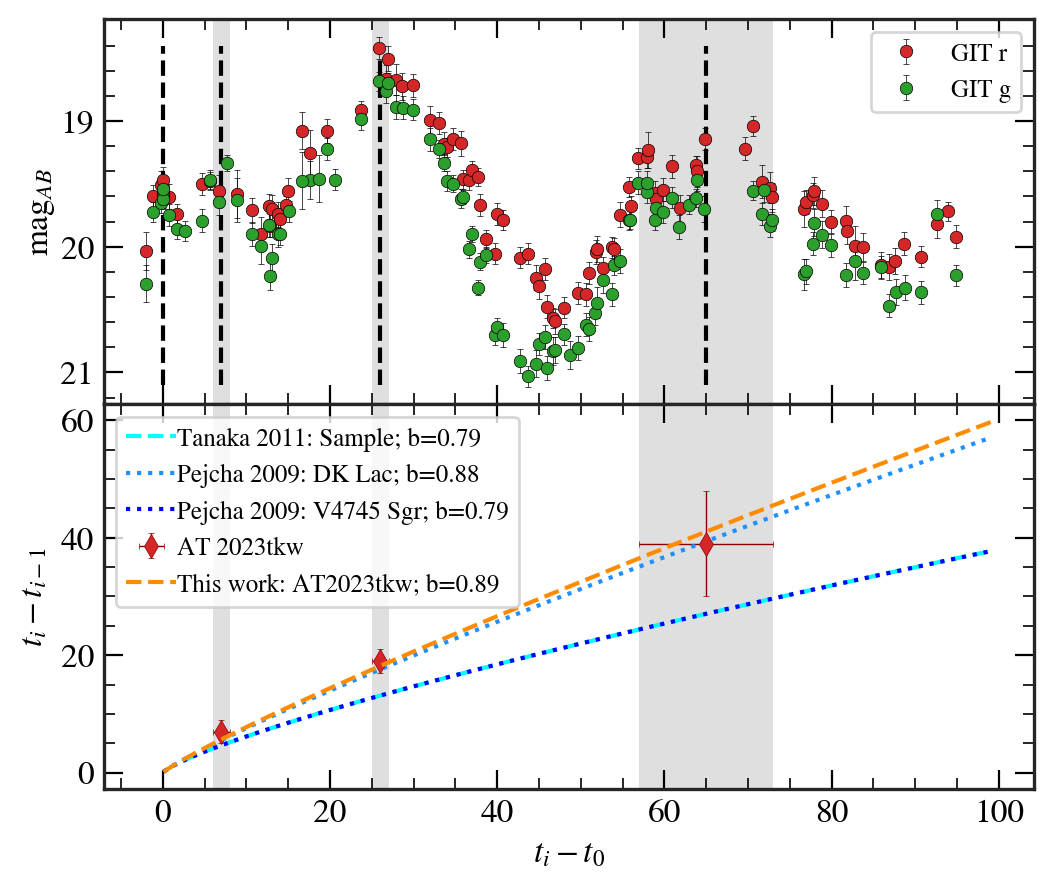}
    \caption{\textit{Top:} Light curve of AT 2023tkw with rebrightening epochs marked with vertical dashed lines, and the errors are shaded. \textit{Bottom:} Time interval between successive maxima plotted against the time from the first maxima. The best fit $\log-\log$ curve is overplotted along with similar values from the literature. The best-fit proportionality constant $b$ is mentioned in the legends for all plots.}
    \label{fig:delt_vs_tmax}
\end{figure}
AT~2023tkw, as discussed in the previous sections, shows explicit modulations in its light curve from its detection till the last observation. The most notable features are the four peaks as shown in Figure~\ref{fig:optical_lc}. Periodogram analysis of these peaks did not reveal any strong signal, suggesting that these features are not due to orbital modulation or any other periodic physical processes. On analyzing the time gap between each successive maximum, we found that it is correlated with the time from the first maxima (see Figure~\ref{fig:delt_vs_tmax}), as was also found by \cite{Pejcha_2009} and \cite{Tanaka_2011} for several novae. The time gap follows the equation below
\begin{equation}
    \log \left( t_{i} - t_{i-1} \right) = b \log \left(t_{i} - t_{0} \right) ; i \in \mathbb{N}
    \label{eqn:delt_vs_tmax}
\end{equation}
where $t_{i}$ is the $i^{th}$ maxima after the first maxima $t_0$. The best-fit value $b=0.89\pm0.12$ was obtained by least square minimization. This value is within the error bars of previously reported results in \cite{Pejcha_2009} and \cite{Tanaka_2011}, and is shown in Figure~\ref{fig:delt_vs_tmax}.

Long oscillatory cycles of rebrightenings were first pointed out by \cite{Bianchini_1992}, suggesting such behaviors could be caused by a balance between gas and radiation pressure when the system goes below the Eddington limit. \cite{Csak_2005} and \cite{Ak_2005} found that the gap between the light curve peaks increases with time linearly for V4745 Sgr and V2540 Oph, respectively. Similar patterns were noted for a sample of novae, including V4745 Sgr and V2540 Oph, by \cite{Tanaka_2011}, who derived an almost linear relationship between successive peaks' intervals and the time since the first maxima (see Figure~\ref{fig:delt_vs_tmax}). \cite{Pejcha_2009} discusses hydrogen burning instabilities as a likely cause for time-dependent rebrightenings and mentions its connection to the decline rate and other yet-to-be-discovered factors. However, \cite{strope_2010} argued strongly against it, stating that the peaks occur randomly. 

\cite{Ak_2005} suggested a possible intermediate-polar (IP) model to explain this trend in V4250 Oph. V4745 Sgr, which also shows multiple peaks near the outburst, was suggested to be an IP by \cite{Dobrotka_2006}. Connections between IPs and oscillatory behavior is not new. \cite{Retter_2002,Retter_2002a} proposed that a lighter disk in IPs is prone to instabilities after the eruption. The disk reformation process is known to be violent, accompanied by disk oscillations. As the disk stabilizes, the rebrightenings become rarer and milder. However, such an explanation is valid only for secondary peaks seen after the primary peak, unlike the case of AT~2023tkw, where we see two rebrightenings before the primary maxima (Figure~\ref{fig:delt_vs_tmax}).

\cite{Steinberg_2020} introduced mass ejection at different time intervals to see the effect of shocks between different layers. These alternate slow and fast outflows could generate multiple correlated $\gamma$-ray and optical peaks. Such peaks were seen to be separated by different time gaps, observed in novae light curves. This suggests that the occurrence of the peaks is linked to the time of origin of the ejected shells within the photosphere.

Such a correlation is seen for the first time in an M31 nova. With more high-cadence surveys of M31, where the novae rates are high, the picture would be clear if this is just a chance event or if there is a causality.

\subsection{Implication of shocks and/or free-free emission on the peak spectrum}
As we noticed in \S \ref{phoenix}, all the spectra of AT~2023tkw show a close resemblance to the model spectra. However, the optical spectrum at the primary peak (top panel in Figure~\ref{fig:spectra_model}) deviates from the \texttt{PHOENIX} atmosphere models. This contradicts the expectation that the emission at the peak is from an expanding photosphere only. It indicates the presence of additional physical processes at the peak of the eruption. As discussed in the previous sections, shocked emission in the UV and X-ray bands can be reprocessed to optical, contributing a significant part to the optical spectrum \citep{Chomiuk_2021}. Due to the lack of $\gamma$-ray observation, we can only speculate the presence of reprocessed shock emission during the peak. Given the correlated emission of $\gamma$-rays and optical found in two recent novae, this is likely the case for many other novae as well. The picture will be more transparent with upcoming simultaneous detections of optical and GeV photons from novae eruptions. 
 
{Another possible explanation for this deviation of observation from the \texttt{PHOENIX} models could be the presence of free-free emission from the ejected winds. Spectra of fast novae are known to be dominated by this free-free emission rather than photospheric emission from an optically thicker region \citep{Hachisu_2006}. However, in slow novae, photospheric contribution plays a significant part (Figure~3 in \citealt{Hachisu_2015}). They show that in the optical passbands, the emission has substantial contributions from both blackbody and optically thick free-free emissions. With time, as the ejecta expands, photospheric emission takes over before recombination and reprocessed radiation from the photoionized ejecta starts to dominate near the dip of the light curve when the ejecta has become optically thin (see bottom panel in Figure~\ref{fig:spectra_model}).


\section{Conclusion}
In this paper, we studied in detail the extragalactic slow classical nova AT~2023tkw in M31 with multiple peaks and dips in its light curve. The object was discovered in the automated transient detection pipeline of the GIT. It was followed up extensively through optical imaging and spectroscopy. The important results from our study are
    \begin{itemize}
    \item The multiple peaks could be caused by shocks that heat the envelope to very high temperatures. The successive peaks seem to occur at longer time intervals caused by internal shocks generated timely within or near the photosphere.
    \item Spectroscopic observations near the peak hint at the presence of another component, apart from the expanding photosphere, contributing to the luminosity. This could be from the shock-heated ionized material.
    \item Spectral modeling using stellar atmosphere codes reveals the expansion and contraction of the photosphere during the rise and decline, respectively, leading to low temperatures at the peak and higher temperatures during the decline. Internal shocks could be responsible for these modulations of the photosphere.
    \item When the ejecta becomes sufficiently optically thin, photoionization starts to dominate. By modeling the spectrum near the dip, we estimated an ejecta mass of $\geq 10^{-4} \mathrm{M_{\odot}}$.
    \item On comparing the observed properties with theoretical models, we derive a low mass WD at 0.65~$\mathrm{M_{\odot}}$ accreting slowly at $5 \times 10^{-13}~\mathrm{M_\odot yr^{-1}}$ from a cool K III giant. This is consistent with slowly evolving classical novae. 
\end{itemize}


\section*{acknowledgement}
The GROWTH-India telescope (GIT) is a 70 cm telescope with a 0.7$^{\circ}$ field of view set up by the Indian Institute of Astrophysics (IIA) and the Indian Institute of Technology Bombay (IITB) with funding from the Indo-US Science and Technology Forum and the Science and Engineering Research Board, Department of Science and Technology, Government of India. It is located at the Indian Astronomical Observatory (IAO, Hanle). We acknowledge funding by the IITB alumni batch of 1994, which partially supports the operation of the telescope. Telescope technical details are available at \url{https://sites.google.com/view/growthindia/}.

Some of the observations reported in this paper were obtained with the 2-m Himalayan Chandra Telescope (HCT). The facilities at IAO and CREST are operated by the Indian Institute of Astrophysics, Bengaluru.

The ZTF forced-photometry service was funded under the Heising-Simons Foundation grant \#12540303 (PI: Graham).

This research is based on observations made with the NASA/ESA Hubble Space Telescope obtained from the Space Telescope Science Institute (STScI/NASA)\footnote{\href{https://archive.stsci.edu/publishing/data-use}{https://archive.stsci.edu/publishing/data-use}}, which is operated by the Association of Universities for Research in Astronomy, Inc., under NASA contract NAS~5–26555. These observations are associated with the program GO-16796.

The HST data described in \S \ref{subsec:hst_phot} is located at the MAST: \dataset[10.17909/b4vg-pa17]{http://dx.doi.org/10.17909/b4vg-pa17}.

This work made use of data supplied by the UK Swift Science Data Centre at the University of Leicester.

G.C. Anupama thanks the Indian National Science Academy for support under the INSA Senior Scientist Programme.

The authors acknowledge the fruitful discussions with Prof Ariel Goobar during the discovery of the object.

The authors thank the referee for their insightful comments and suggestions which improved the content of the paper.

\facilities{\textit{GIT},
            \textit{HCT},
            \textit{HST},
            \textit{Keck:I},
            \textit{Keck:II},
            \textit{Hale},
            \textit{PO:1.2m}
            }
            
\software{  \texttt{IRAF v2.16.1} \citep{tod93}, 
            \texttt{Python v3.6.6} \citep{python09},
            \texttt{NumPy} \citep{NumPy20}, 
            \texttt{SciPy} \citep{SciPy20}, 
            \texttt{Astropy} \citep{astropy:2013, astropy:2018, astropy:2022},
            \texttt{Pandas} \citep{mckinney10}, 
            \texttt{Matplotlib} \citep{Hunter07},
            \texttt{Spalipy} \citep{spalipy},
            \texttt{extinction} \citep{Barbary_2021},
            \texttt{photutils} \citep{Larry_2023},
            \texttt{SWarp} \citep{swarp},
            \texttt{SAOImageDS9} \citep{ds9_2003},
            \texttt{SExtractor } \citep{ Bertin_1996}
            }

\bibliography{main}{}

\begin{thebibliography}{}
\expandafter\ifx\csname natexlab\endcsname\relax\def\natexlab#1{#1}\fi
\providecommand{\url}[1]{\href{#1}{#1}}
\providecommand{\dodoi}[1]{doi:~\href{http://doi.org/#1}{\nolinkurl{#1}}}
\providecommand{\doeprint}[1]{\href{http://ascl.net/#1}{\nolinkurl{http://ascl.net/#1}}}
\providecommand{\doarXiv}[1]{\href{https://arxiv.org/abs/#1}{\nolinkurl{https://arxiv.org/abs/#1}}}

\bibitem[{{Ak} {et~al.}(2005){Ak}, {Retter}, \& {Liu}}]{Ak_2005}
{Ak}, T., {Retter}, A., \& {Liu}, A. 2005, \pasa, 22, 298, \dodoi{10.1071/AS05017}

\bibitem[{{Anupama} {et~al.}(2013){Anupama}, {Kamath}, {Ramaprakash}, {Kantharia}, {Hegde}, {Mohan}, {Kulkarni}, {Bode}, {Eyres}, {Evans}, \& {O'Brien}}]{Anupama_2013}
{Anupama}, G.~C., {Kamath}, U.~S., {Ramaprakash}, A.~N., {et~al.} 2013, \aap, 559, A121, \dodoi{10.1051/0004-6361/201321262}

\bibitem[{{Astropy Collaboration} {et~al.}(2013){Astropy Collaboration}, {Robitaille}, {Tollerud}, {Greenfield}, {Droettboom}, {Bray}, {Aldcroft}, {Davis}, {Ginsburg}, {Price-Whelan}, {Kerzendorf}, {Conley}, {Crighton}, {Barbary}, {Muna}, {Ferguson}, {Grollier}, {Parikh}, {Nair}, {Unther}, {Deil}, {Woillez}, {Conseil}, {Kramer}, {Turner}, {Singer}, {Fox}, {Weaver}, {Zabalza}, {Edwards}, {Azalee Bostroem}, {Burke}, {Casey}, {Crawford}, {Dencheva}, {Ely}, {Jenness}, {Labrie}, {Lim}, {Pierfederici}, {Pontzen}, {Ptak}, {Refsdal}, {Servillat}, \& {Streicher}}]{astropy:2013}
{Astropy Collaboration}, {Robitaille}, T.~P., {Tollerud}, E.~J., {et~al.} 2013, \aap, 558, A33, \dodoi{10.1051/0004-6361/201322068}

\bibitem[{{Astropy Collaboration} {et~al.}(2018){Astropy Collaboration}, {Price-Whelan}, {Sip{\H{o}}cz}, {G{\"u}nther}, {Lim}, {Crawford}, {Conseil}, {Shupe}, {Craig}, {Dencheva}, {Ginsburg}, {Vand erPlas}, {Bradley}, {P{\'e}rez-Su{\'a}rez}, {de Val-Borro}, {Aldcroft}, {Cruz}, {Robitaille}, {Tollerud}, {Ardelean}, {Babej}, {Bach}, {Bachetti}, {Bakanov}, {Bamford}, {Barentsen}, {Barmby}, {Baumbach}, {Berry}, {Biscani}, {Boquien}, {Bostroem}, {Bouma}, {Brammer}, {Bray}, {Breytenbach}, {Buddelmeijer}, {Burke}, {Calderone}, {Cano Rodr{\'\i}guez}, {Cara}, {Cardoso}, {Cheedella}, {Copin}, {Corrales}, {Crichton}, {D'Avella}, {Deil}, {Depagne}, {Dietrich}, {Donath}, {Droettboom}, {Earl}, {Erben}, {Fabbro}, {Ferreira}, {Finethy}, {Fox}, {Garrison}, {Gibbons}, {Goldstein}, {Gommers}, {Greco}, {Greenfield}, {Groener}, {Grollier}, {Hagen}, {Hirst}, {Homeier}, {Horton}, {Hosseinzadeh}, {Hu}, {Hunkeler}, {Ivezi{\'c}}, {Jain}, {Jenness}, {Kanarek}, {Kendrew}, {Kern}, {Kerzendorf}, {Khvalko}, {King}, {Kirkby}, {Kulkarni},
  {Kumar}, {Lee}, {Lenz}, {Littlefair}, {Ma}, {Macleod}, {Mastropietro}, {McCully}, {Montagnac}, {Morris}, {Mueller}, {Mumford}, {Muna}, {Murphy}, {Nelson}, {Nguyen}, {Ninan}, {N{\"o}the}, {Ogaz}, {Oh}, {Parejko}, {Parley}, {Pascual}, {Patil}, {Patil}, {Plunkett}, {Prochaska}, {Rastogi}, {Reddy Janga}, {Sabater}, {Sakurikar}, {Seifert}, {Sherbert}, {Sherwood-Taylor}, {Shih}, {Sick}, {Silbiger}, {Singanamalla}, {Singer}, {Sladen}, {Sooley}, {Sornarajah}, {Streicher}, {Teuben}, {Thomas}, {Tremblay}, {Turner}, {Terr{\'o}n}, {van Kerkwijk}, {de la Vega}, {Watkins}, {Weaver}, {Whitmore}, {Woillez}, {Zabalza}, \& {Astropy Contributors}}]{astropy:2018}
{Astropy Collaboration}, {Price-Whelan}, A.~M., {Sip{\H{o}}cz}, B.~M., {et~al.} 2018, \aj, 156, 123, \dodoi{10.3847/1538-3881/aabc4f}

\bibitem[{{Astropy Collaboration} {et~al.}(2022){Astropy Collaboration}, {Price-Whelan}, {Lim}, {Earl}, {Starkman}, {Bradley}, {Shupe}, {Patil}, {Corrales}, {Brasseur}, {N{"o}the}, {Donath}, {Tollerud}, {Morris}, {Ginsburg}, {Vaher}, {Weaver}, {Tocknell}, {Jamieson}, {van Kerkwijk}, {Robitaille}, {Merry}, {Bachetti}, {G{"u}nther}, {Aldcroft}, {Alvarado-Montes}, {Archibald}, {B{'o}di}, {Bapat}, {Barentsen}, {Baz{'a}n}, {Biswas}, {Boquien}, {Burke}, {Cara}, {Cara}, {Conroy}, {Conseil}, {Craig}, {Cross}, {Cruz}, {D'Eugenio}, {Dencheva}, {Devillepoix}, {Dietrich}, {Eigenbrot}, {Erben}, {Ferreira}, {Foreman-Mackey}, {Fox}, {Freij}, {Garg}, {Geda}, {Glattly}, {Gondhalekar}, {Gordon}, {Grant}, {Greenfield}, {Groener}, {Guest}, {Gurovich}, {Handberg}, {Hart}, {Hatfield-Dodds}, {Homeier}, {Hosseinzadeh}, {Jenness}, {Jones}, {Joseph}, {Kalmbach}, {Karamehmetoglu}, {Ka{l}uszy{'n}ski}, {Kelley}, {Kern}, {Kerzendorf}, {Koch}, {Kulumani}, {Lee}, {Ly}, {Ma}, {MacBride}, {Maljaars}, {Muna}, {Murphy}, {Norman}, {O'Steen},
  {Oman}, {Pacifici}, {Pascual}, {Pascual-Granado}, {Patil}, {Perren}, {Pickering}, {Rastogi}, {Roulston}, {Ryan}, {Rykoff}, {Sabater}, {Sakurikar}, {Salgado}, {Sanghi}, {Saunders}, {Savchenko}, {Schwardt}, {Seifert-Eckert}, {Shih}, {Jain}, {Shukla}, {Sick}, {Simpson}, {Singanamalla}, {Singer}, {Singhal}, {Sinha}, {Sip{H{o}}cz}, {Spitler}, {Stansby}, {Streicher}, {{{S}}umak}, {Swinbank}, {Taranu}, {Tewary}, {Tremblay}, {Val-Borro}, {Van Kooten}, {Vasovi{'c}}, {Verma}, {de Miranda Cardoso}, {Williams}, {Wilson}, {Winkel}, {Wood-Vasey}, {Xue}, {Yoachim}, {Zhang}, {Zonca}, \& {Astropy Project Contributors}}]{astropy:2022}
{Astropy Collaboration}, {Price-Whelan}, A.~M., {Lim}, P.~L., {et~al.} 2022, \apj, 935, 167, \dodoi{10.3847/1538-4357/ac7c74}

\bibitem[{{Aydi} {et~al.}(2020){Aydi}, {Sokolovsky}, {Chomiuk}, {Steinberg}, {Li}, {Vurm}, {Metzger}, {Strader}, {Mukai}, {Pejcha}, {Shen}, {Wade}, {Kuschnig}, {Moffat}, {Pablo}, {Pigulski}, {Popowicz}, {Weiss}, {Zwintz}, {Izzo}, {Pollard}, {Handler}, {Ryder}, {Filipovi{\'c}}, {Alsaberi}, {Manojlovi{\'c}}, {Lopes de Oliveira}, {Walter}, {Vallely}, {Buckley}, {Brown}, {Harvey}, {Kawash}, {Kniazev}, {Kochanek}, {Linford}, {Mikolajewska}, {Molaro}, {Orio}, {Page}, {Shappee}, \& {Sokoloski}}]{Aydi_2020a}
{Aydi}, E., {Sokolovsky}, K.~V., {Chomiuk}, L., {et~al.} 2020, Nature Astronomy, 4, 776, \dodoi{10.1038/s41550-020-1070-y}

\bibitem[{{Aydi} {et~al.}(2023){Aydi}, {Chomiuk}, {Miko{\l}ajewska}, {Brink}, {Metzger}, {Strader}, {Buckley}, {Harvey}, {Holoien}, {Izzo}, {Kawash}, {Linford}, {Molaro}, {Molina}, {Mr{\'o}z}, {Mukai}, {Orio}, {Panurach}, {Senchyna}, {Shappee}, {Shen}, {Sokoloski}, {Sokolovsky}, {Urquhart}, \& {Williams}}]{Aydi_2023}
{Aydi}, E., {Chomiuk}, L., {Miko{\l}ajewska}, J., {et~al.} 2023, \mnras, 524, 1946, \dodoi{10.1093/mnras/stad1914}

\bibitem[{{Aydi} {et~al.}(2024){Aydi}, {Chomiuk}, {Strader}, {Sokolovsky}, {Williams}, {Buckley}, {Ederoclite}, {Izzo}, {Kyer}, {Linford}, {Kniazev}, {Metzger}, {Miko{\l}ajewska}, {Molaro}, {Molina}, {Mukai}, {Munari}, {Orio}, {Panurach}, {Shappee}, {Shen}, {Sokoloski}, {Urquhart}, \& {Walter}}]{Aydi_2024}
{Aydi}, E., {Chomiuk}, L., {Strader}, J., {et~al.} 2024, \mnras, 527, 9303, \dodoi{10.1093/mnras/stad3342}

\bibitem[{{Banerjee} \& {Ashok}(2012)}]{banerjee_2012}
{Banerjee}, D.~P.~K., \& {Ashok}, N.~M. 2012, Bulletin of the Astronomical Society of India, 40, 243, \dodoi{10.48550/arXiv.1306.0343}

\bibitem[{{Banerjee} {et~al.}(2010){Banerjee}, {Das}, {Ashok}, {Rushton}, {Eyres}, {Maxwell}, {Worters}, {Evans}, \& {Schaefer}}]{2010MNRAS.408L..71B}
{Banerjee}, D.~P.~K., {Das}, R.~K., {Ashok}, N.~M., {et~al.} 2010, \mnras, 408, L71, \dodoi{10.1111/j.1745-3933.2010.00932.x}

\bibitem[{{Barbary}(2021)}]{Barbary_2021}
{Barbary}, K. 2021, {extinction: Dust extinction laws}, Astrophysics Source Code Library, record ascl:2102.026

\bibitem[{{Basu} {et~al.}(2024{\natexlab{a}}){Basu}, {Krishnendu}, {Barway}, {Chamoli}, \& {Anupama}}]{basu_2024a}
{Basu}, J., {Krishnendu}, S., {Barway}, S., {Chamoli}, S., \& {Anupama}, G.~C. 2024{\natexlab{a}}, \apj, 971, 8, \dodoi{10.3847/1538-4357/ad596b}

\bibitem[{{Basu} {et~al.}(2023){Basu}, {Kumar}, {Anupama}, {Barway}, {Bhalerao}, {Swain}, {Singh}, {Teja}, {Kurre}, {Patel}, \& {Bandari}}]{Basu_2023}
{Basu}, J., {Kumar}, R., {Anupama}, G.~C., {et~al.} 2023, The Astronomer's Telegram, 16311, 1

\bibitem[{{Basu} {et~al.}(2024{\natexlab{b}}){Basu}, {Pavana}, {Anupama}, {Barway}, {Singh}, {Swain}, {Srivastav}, {Kumar}, {Bhalerao}, {Sonith}, \& {Selvakumar}}]{basu_2024}
{Basu}, J., {Pavana}, M., {Anupama}, G.~C., {et~al.} 2024{\natexlab{b}}, \apj, 966, 44, \dodoi{10.3847/1538-4357/ad2c8e}

\bibitem[{{Beardmore} {et~al.}(2012){Beardmore}, {Osborne}, {Page}, {Hakala}, {Schwarz}, {Rauch}, {Balman}, {Evans}, {Goad}, {Ness}, {Starrfield}, \& {Wagner}}]{Beardmore_2012}
{Beardmore}, A.~P., {Osborne}, J.~P., {Page}, K.~L., {et~al.} 2012, \aap, 545, A116, \dodoi{10.1051/0004-6361/201219681}

\bibitem[{{Bellm} {et~al.}(2019){Bellm}, {Kulkarni}, {Graham}, {Dekany}, {Smith}, {Riddle}, {Masci}, {Helou}, {Prince}, {Adams}, {Barbarino}, {Barlow}, {Bauer}, {Beck}, {Belicki}, {Biswas}, {Blagorodnova}, {Bodewits}, {Bolin}, {Brinnel}, {Brooke}, {Bue}, {Bulla}, {Burruss}, {Cenko}, {Chang}, {Connolly}, {Coughlin}, {Cromer}, {Cunningham}, {De}, {Delacroix}, {Desai}, {Duev}, {Eadie}, {Farnham}, {Feeney}, {Feindt}, {Flynn}, {Franckowiak}, {Frederick}, {Fremling}, {Gal-Yam}, {Gezari}, {Giomi}, {Goldstein}, {Golkhou}, {Goobar}, {Groom}, {Hacopians}, {Hale}, {Henning}, {Ho}, {Hover}, {Howell}, {Hung}, {Huppenkothen}, {Imel}, {Ip}, {Ivezi{\'c}}, {Jackson}, {Jones}, {Juric}, {Kasliwal}, {Kaspi}, {Kaye}, {Kelley}, {Kowalski}, {Kramer}, {Kupfer}, {Landry}, {Laher}, {Lee}, {Lin}, {Lin}, {Lunnan}, {Giomi}, {Mahabal}, {Mao}, {Miller}, {Monkewitz}, {Murphy}, {Ngeow}, {Nordin}, {Nugent}, {Ofek}, {Patterson}, {Penprase}, {Porter}, {Rauch}, {Rebbapragada}, {Reiley}, {Rigault}, {Rodriguez}, {van Roestel}, {Rusholme}, {van
  Santen}, {Schulze}, {Shupe}, {Singer}, {Soumagnac}, {Stein}, {Surace}, {Sollerman}, {Szkody}, {Taddia}, {Terek}, {Van Sistine}, {van Velzen}, {Vestrand}, {Walters}, {Ward}, {Ye}, {Yu}, {Yan}, \& {Zolkower}}]{Bellm_2019}
{Bellm}, E.~C., {Kulkarni}, S.~R., {Graham}, M.~J., {et~al.} 2019, \pasp, 131, 018002, \dodoi{10.1088/1538-3873/aaecbe}

\bibitem[{{Bertin}(2010)}]{swarp}
{Bertin}, E. 2010, {SWarp: Resampling and Co-adding FITS Images Together}, Astrophysics Source Code Library, record ascl:1010.068

\bibitem[{{Bertin} \& {Arnouts}(1996)}]{Bertin_1996}
{Bertin}, E., \& {Arnouts}, S. 1996, \aaps, 117, 393, \dodoi{10.1051/aas:1996164}

\bibitem[{{Bianchini} {et~al.}(1992){Bianchini}, {Friedjung}, \& {Brinkmann}}]{Bianchini_1992}
{Bianchini}, A., {Friedjung}, M., \& {Brinkmann}, W. 1992, \aap, 257, 599

\bibitem[{{Bode} \& {Evans}(2008)}]{2008clno.book.....B}
{Bode}, M.~F., \& {Evans}, A. 2008, {Classical Novae}, Vol.~43, \dodoi{10.1017/CBO9780511536168}

\bibitem[{Bradley {et~al.}(2023)Bradley, Sip{\H o}cz, Robitaille, Tollerud, Vin{\'{\i}}cius, Deil, Barbary, Wilson, Busko, Donath, G{\"u}nther, Cara, Lim, Me{\ss}linger, Conseil, Bostroem, Droettboom, Bray, Bratholm, Barentsen, Craig, Rathi, Pascual, Perren, Georgiev, de~Val-Borro, Kerzendorf, Bach, Quint, \& Souchereau}]{Larry_2023}
Bradley, L., Sip{\H o}cz, B., Robitaille, T., {et~al.} 2023, astropy/photutils: 1.8.0, 1.8.0,  Zenodo, \dodoi{10.5281/zenodo.7946442}

\bibitem[{{Buson} {et~al.}(2021){Buson}, {Cheung}, \& {Jean}}]{Buson_2021}
{Buson}, S., {Cheung}, C.~C., \& {Jean}, P. 2021, The Astronomer's Telegram, 14658, 1

\bibitem[{{Cao} {et~al.}(2012){Cao}, {Kasliwal}, {Neill}, {Kulkarni}, {Lou}, {Ben-Ami}, {Bloom}, {Cenko}, {Law}, {Nugent}, {Ofek}, {Poznanski}, \& {Quimby}}]{Cao_2012}
{Cao}, Y., {Kasliwal}, M.~M., {Neill}, J.~D., {et~al.} 2012, \apj, 752, 133, \dodoi{10.1088/0004-637X/752/2/133}

\bibitem[{{Cardelli} {et~al.}(1989){Cardelli}, {Clayton}, \& {Mathis}}]{Cardelli_1989}
{Cardelli}, J.~A., {Clayton}, G.~C., \& {Mathis}, J.~S. 1989, \apj, 345, 245, \dodoi{10.1086/167900}

\bibitem[{{Chambers} {et~al.}(2016){Chambers}, {Magnier}, {Metcalfe}, {Flewelling}, {Huber}, {Waters}, {Denneau}, {Draper}, {Farrow}, {Finkbeiner}, {Holmberg}, {Koppenhoefer}, {Price}, {Rest}, {Saglia}, {Schlafly}, {Smartt}, {Sweeney}, {Wainscoat}, {Burgett}, {Chastel}, {Grav}, {Heasley}, {Hodapp}, {Jedicke}, {Kaiser}, {Kudritzki}, {Luppino}, {Lupton}, {Monet}, {Morgan}, {Onaka}, {Shiao}, {Stubbs}, {Tonry}, {White}, {Ba{\~n}ados}, {Bell}, {Bender}, {Bernard}, {Boegner}, {Boffi}, {Botticella}, {Calamida}, {Casertano}, {Chen}, {Chen}, {Cole}, {Deacon}, {Frenk}, {Fitzsimmons}, {Gezari}, {Gibbs}, {Goessl}, {Goggia}, {Gourgue}, {Goldman}, {Grant}, {Grebel}, {Hambly}, {Hasinger}, {Heavens}, {Heckman}, {Henderson}, {Henning}, {Holman}, {Hopp}, {Ip}, {Isani}, {Jackson}, {Keyes}, {Koekemoer}, {Kotak}, {Le}, {Liska}, {Long}, {Lucey}, {Liu}, {Martin}, {Masci}, {McLean}, {Mindel}, {Misra}, {Morganson}, {Murphy}, {Obaika}, {Narayan}, {Nieto-Santisteban}, {Norberg}, {Peacock}, {Pier}, {Postman}, {Primak}, {Rae}, {Rai},
  {Riess}, {Riffeser}, {Rix}, {R{\"o}ser}, {Russel}, {Rutz}, {Schilbach}, {Schultz}, {Scolnic}, {Strolger}, {Szalay}, {Seitz}, {Small}, {Smith}, {Soderblom}, {Taylor}, {Thomson}, {Taylor}, {Thakar}, {Thiel}, {Thilker}, {Unger}, {Urata}, {Valenti}, {Wagner}, {Walder}, {Walter}, {Watters}, {Werner}, {Wood-Vasey}, \& {Wyse}}]{chambers_2016}
{Chambers}, K.~C., {Magnier}, E.~A., {Metcalfe}, N., {et~al.} 2016, arXiv e-prints, arXiv:1612.05560, \dodoi{10.48550/arXiv.1612.05560}

\bibitem[{{Chatzikos} {et~al.}(2023){Chatzikos}, {Bianchi}, {Camilloni}, {Chakraborty}, {Gunasekera}, {Guzm{\'a}n}, {Milby}, {Sarkar}, {Shaw}, {van Hoof}, \& {Ferland}}]{cloudy23}
{Chatzikos}, M., {Bianchi}, S., {Camilloni}, F., {et~al.} 2023, \rmxaa, 59, 327, \dodoi{10.22201/ia.01851101p.2023.59.02.12}

\bibitem[{{Cheung}(2024)}]{Cheung_2024}
{Cheung}, C.~C. 2024, The Astronomer's Telegram, 16439, 1

\bibitem[{{Cheung} \& {Jean}(2024)}]{Cheung_2024b}
{Cheung}, C.~C., \& {Jean}, P. 2024, The Astronomer's Telegram, 16441, 1

\bibitem[{{Cheung} {et~al.}(2016){Cheung}, {Jean}, {Shore}, {Stawarz}, {Corbet}, {Kn{\"o}dlseder}, {Starrfield}, {Wood}, {Desiante}, {Longo}, {Pivato}, \& {Wood}}]{Cheung_2016}
{Cheung}, C.~C., {Jean}, P., {Shore}, S.~N., {et~al.} 2016, \apj, 826, 142, \dodoi{10.3847/0004-637X/826/2/142}

\bibitem[{{Chomiuk} {et~al.}(2021){Chomiuk}, {Metzger}, \& {Shen}}]{Chomiuk_2021}
{Chomiuk}, L., {Metzger}, B.~D., \& {Shen}, K.~J. 2021, \araa, 59, 391, \dodoi{10.1146/annurev-astro-112420-114502}

\bibitem[{{Clark} {et~al.}(2024){Clark}, {Hornoch}, {Shafter}, {Ku{\v{c}}{\'a}kov{\'a}}, {Vra{\v{s}}til}, {Ku{\v{s}}nir{\'a}k}, \& {Wolf}}]{Clark_2024}
{Clark}, J.~G., {Hornoch}, K., {Shafter}, A.~W., {et~al.} 2024, \apjs, 272, 28, \dodoi{10.3847/1538-4365/ad3c39}

\bibitem[{{Cs{\'a}k} {et~al.}(2005){Cs{\'a}k}, {Kiss}, {Retter}, {Jacob}, \& {Kaspi}}]{Csak_2005}
{Cs{\'a}k}, B., {Kiss}, L.~L., {Retter}, A., {Jacob}, A., \& {Kaspi}, S. 2005, \aap, 429, 599, \dodoi{10.1051/0004-6361:20035751}

\bibitem[{{Cushing} {et~al.}(2004){Cushing}, {Vacca}, \& {Rayner}}]{Cushing_2004}
{Cushing}, M.~C., {Vacca}, W.~D., \& {Rayner}, J.~T. 2004, \pasp, 116, 362, \dodoi{10.1086/382907}

\bibitem[{{Darnley} \& {Henze}(2020)}]{Darnley_2020}
{Darnley}, M.~J., \& {Henze}, M. 2020, Advances in Space Research, 66, 1147, \dodoi{10.1016/j.asr.2019.09.044}

\bibitem[{{Darnley} {et~al.}(2006){Darnley}, {Bode}, {Kerins}, {Newsam}, {An}, {Baillon}, {Belokurov}, {Calchi Novati}, {Carr}, {Cr{\'e}z{\'e}}, {Evans}, {Giraud-H{\'e}raud}, {Gould}, {Hewett}, {Jetzer}, {Kaplan}, {Paulin-Henriksson}, {Smartt}, {Tsapras}, \& {Weston}}]{Darnley_2006}
{Darnley}, M.~J., {Bode}, M.~F., {Kerins}, E., {et~al.} 2006, \mnras, 369, 257, \dodoi{10.1111/j.1365-2966.2006.10297.x}

\bibitem[{Dekany {et~al.}(2020)Dekany, Smith, Riddle, Feeney, Porter, Hale, Zolkower, Belicki, Kaye, Henning, Walters, Cromer, Delacroix, Rodriguez, Reiley, Mao, Hover, Murphy, Burruss, Baker, Kowalski, Reif, Mueller, Bellm, Graham, \& Kulkarni}]{Dekany_2020}
Dekany, R., Smith, R.~M., Riddle, R., {et~al.} 2020, Publications of the Astronomical Society of the Pacific, 132, 038001, \dodoi{10.1088/1538-3873/ab4ca2}

\bibitem[{{Dobrotka} {et~al.}(2006){Dobrotka}, {Retter}, \& {Liu}}]{Dobrotka_2006}
{Dobrotka}, A., {Retter}, A., \& {Liu}, A. 2006, \mnras, 371, 459, \dodoi{10.1111/j.1365-2966.2006.10683.x}

\bibitem[{{Dolphin}(2016)}]{dolphot}
{Dolphin}, A. 2016, {DOLPHOT: Stellar photometry}, Astrophysics Source Code Library, record ascl:1608.013

\bibitem[{{Dolphin}(2000)}]{hstphot}
{Dolphin}, A.~E. 2000, \pasp, 112, 1383, \dodoi{10.1086/316630}

\bibitem[{{Draine} {et~al.}(2014){Draine}, {Aniano}, {Krause}, {Groves}, {Sandstrom}, {Braun}, {Leroy}, {Klaas}, {Linz}, {Rix}, {Schinnerer}, {Schmiedeke}, \& {Walter}}]{Draine_2014}
{Draine}, B.~T., {Aniano}, G., {Krause}, O., {et~al.} 2014, \apj, 780, 172, \dodoi{10.1088/0004-637X/780/2/172}

\bibitem[{{Evans} {et~al.}(2003){Evans}, {Gehrz}, {Geballe}, {Woodward}, {Salama}, {Sanchez}, {Starrfield}, {Krautter}, {Barlow}, {Lyke}, {Hayward}, {Eyres}, {Greenhouse}, {Hjellming}, {Wagner}, \& {Pequignot}}]{Evans_2003}
{Evans}, A., {Gehrz}, R.~D., {Geballe}, T.~R., {et~al.} 2003, \aj, 126, 1981, \dodoi{10.1086/377618}

\bibitem[{Evans {et~al.}(2020)Evans, Page, Osborne, Beardmore, Willingale, Burrows, Kennea, Perri, Capalbi, Tagliaferri, \& Cenko}]{Evans_2020}
Evans, P.~A., Page, K.~L., Osborne, J.~P., {et~al.} 2020, The Astrophysical Journal Supplement Series, 247, 54, \dodoi{10.3847/1538-4365/ab7db9}

\bibitem[{{Ferland} {et~al.}(2017){Ferland}, {Chatzikos}, {Guzm{\'a}n}, {Lykins}, {van Hoof}, {Williams}, {Abel}, {Badnell}, {Keenan}, {Porter}, \& {Stancil}}]{cloudy17}
{Ferland}, G.~J., {Chatzikos}, M., {Guzm{\'a}n}, F., {et~al.} 2017, \rmxaa, 53, 385, \dodoi{10.48550/arXiv.1705.10877}

\bibitem[{{Fuentes-Morales} {et~al.}(2021){Fuentes-Morales}, {Tappert}, {Zorotovic}, {Vogt}, {Puebla}, {Schreiber}, {Ederoclite}, \& {Schmidtobreick}}]{Morales_2021}
{Fuentes-Morales}, I., {Tappert}, C., {Zorotovic}, M., {et~al.} 2021, \mnras, 501, 6083, \dodoi{10.1093/mnras/staa3482}

\bibitem[{{Gehrz} {et~al.}(1998){Gehrz}, {Truran}, {Williams}, \& {Starrfield}}]{Gehrz_1998}
{Gehrz}, R.~D., {Truran}, J.~W., {Williams}, R.~E., \& {Starrfield}, S. 1998, \pasp, 110, 3, \dodoi{10.1086/316107}

\bibitem[{{Gordon} {et~al.}(2021){Gordon}, {Aydi}, {Page}, {Li}, {Chomiuk}, {Sokolovsky}, {Mukai}, \& {Seitz}}]{Gordon_2021}
{Gordon}, A.~C., {Aydi}, E., {Page}, K.~L., {et~al.} 2021, \apj, 910, 134, \dodoi{10.3847/1538-4357/abe547}

\bibitem[{Graham {et~al.}(2019)Graham, Kulkarni, Bellm, Adams, Barbarino, Blagorodnova, Bodewits, Bolin, Brady, Cenko, Chang, Coughlin, De, Eadie, Farnham, Feindt, Franckowiak, Fremling, Gezari, Ghosh, Goldstein, Golkhou, Goobar, Ho, Huppenkothen, Željko Ivezić, Jones, Juric, Kaplan, Kasliwal, Kelley, Kupfer, Lee, Lin, Lunnan, Mahabal, Miller, Ngeow, Nugent, Ofek, Prince, Rauch, van Roestel, Schulze, Singer, Sollerman, Taddia, Yan, Ye, Yu, Barlow, Bauer, Beck, Belicki, Biswas, Brinnel, Brooke, Bue, Bulla, Burruss, Connolly, Cromer, Cunningham, Dekany, Delacroix, Desai, Duev, Feeney, Flynn, Frederick, Gal-Yam, Giomi, Groom, Hacopians, Hale, Helou, Henning, Hover, Hillenbrand, Howell, Hung, Imel, Ip, Jackson, Kaspi, Kaye, Kowalski, Kramer, Kuhn, Landry, Laher, Mao, Masci, Monkewitz, Murphy, Nordin, Patterson, Penprase, Porter, Rebbapragada, Reiley, Riddle, Rigault, Rodriguez, Rusholme, van Santen, Shupe, Smith, Soumagnac, Stein, Surace, Szkody, Terek, Sistine, van Velzen, Vestrand, Walters, Ward, Zhang, \&
  Zolkower}]{Graham_2019}
Graham, M.~J., Kulkarni, S.~R., Bellm, E.~C., {et~al.} 2019, Publications of the Astronomical Society of the Pacific, 131, 078001, \dodoi{10.1088/1538-3873/ab006c}

\bibitem[{{Grevesse} {et~al.}(2010){Grevesse}, {Asplund}, {Sauval}, \& {Scott}}]{GASS10}
{Grevesse}, N., {Asplund}, M., {Sauval}, A.~J., \& {Scott}, P. 2010, \apss, 328, 179, \dodoi{10.1007/s10509-010-0288-z}

\bibitem[{{Hachisu} \& {Kato}(2006)}]{Hachisu_2006}
{Hachisu}, I., \& {Kato}, M. 2006, \apjs, 167, 59, \dodoi{10.1086/508063}

\bibitem[{{Hachisu} \& {Kato}(2015)}]{Hachisu_2015}
---. 2015, \apj, 798, 76, \dodoi{10.1088/0004-637X/798/2/76}

\bibitem[{Harris {et~al.}(2020)Harris, Millman, van~der Walt, Gommers, Virtanen, Cournapeau, Wieser, Taylor, Berg, Smith, Kern, Picus, Hoyer, van Kerkwijk, Brett, Haldane, Fernández~del Río, Wiebe, Peterson, Gérard-Marchant, Sheppard, Reddy, Weckesser, Abbasi, Gohlke, \& Oliphant}]{NumPy20}
Harris, C.~R., Millman, K.~J., van~der Walt, S.~J., {et~al.} 2020, Nature, 585, 357–362, \dodoi{10.1038/s41586-020-2649-2}

\bibitem[{{Harris}(2018)}]{Harris_2018}
{Harris}, W.~E. 2018, \aj, 156, 296, \dodoi{10.3847/1538-3881/aaedb8}

\bibitem[{{Hathi}(2024)}]{Hathi_2024}
{Hathi}, N.~P. 2024, in ACS Data Handbook v. 13.0, Vol.~13, 13

\bibitem[{{Hauschildt} \& {Baron}(1995)}]{Hauschildt_1995}
{Hauschildt}, P.~H., \& {Baron}, E. 1995, \jqsrt, 54, 987, \dodoi{10.1016/0022-4073(95)00118-5}

\bibitem[{{Hauschildt} {et~al.}(1997){Hauschildt}, {Shore}, {Schwarz}, {Baron}, {Starrfield}, \& {Allard}}]{Hauschildt_1997}
{Hauschildt}, P.~H., {Shore}, S.~N., {Schwarz}, G.~J., {et~al.} 1997, \apj, 490, 803, \dodoi{10.1086/304904}

\bibitem[{{Hillman} {et~al.}(2016){Hillman}, {Prialnik}, {Kovetz}, \& {Shara}}]{Hillman_2016}
{Hillman}, Y., {Prialnik}, D., {Kovetz}, A., \& {Shara}, M.~M. 2016, \apj, 819, 168, \dodoi{10.3847/0004-637X/819/2/168}

\bibitem[{Hunter(2007)}]{Hunter07}
Hunter, J.~D. 2007, Computing in Science \& Engineering, 9, 90, \dodoi{10.1109/MCSE.2007.55}

\bibitem[{{Husser} {et~al.}(2013){Husser}, {Wende-von Berg}, {Dreizler}, {Homeier}, {Reiners}, {Barman}, \& {Hauschildt}}]{Husser_2013}
{Husser}, T.~O., {Wende-von Berg}, S., {Dreizler}, S., {et~al.} 2013, \aap, 553, A6, \dodoi{10.1051/0004-6361/201219058}

\bibitem[{{Joye} \& {Mandel}(2003)}]{ds9_2003}
{Joye}, W.~A., \& {Mandel}, E. 2003, in Astronomical Society of the Pacific Conference Series, Vol. 295, Astronomical Data Analysis Software and Systems XII, ed. H.~E. {Payne}, R.~I. {Jedrzejewski}, \& R.~N. {Hook}, 489

\bibitem[{{Kastner} \& {Bhatia}(1995)}]{Kastner_1995}
{Kastner}, S.~O., \& {Bhatia}, A.~K. 1995, \apj, 439, 346, \dodoi{10.1086/175178}

\bibitem[{{Kato} \& {Hachisu}(2009)}]{Kato_2009}
{Kato}, M., \& {Hachisu}, I. 2009, \apj, 699, 1293, \dodoi{10.1088/0004-637X/699/2/1293}

\bibitem[{{Kato} \& {Hachisu}(2011)}]{kato_2011}
---. 2011, \apj, 743, 157, \dodoi{10.1088/0004-637X/743/2/157}

\bibitem[{{Kato} {et~al.}(2022){Kato}, {Saio}, \& {Hachisu}}]{Kato_2022}
{Kato}, M., {Saio}, H., \& {Hachisu}, I. 2022, \pasj, 74, 1005, \dodoi{10.1093/pasj/psac051}

\bibitem[{{Kato} {et~al.}(2024){Kato}, {Saio}, \& {Hachisu}}]{Kato_2024}
---. 2024, \pasj, 76, 666, \dodoi{10.1093/pasj/psae038}

\bibitem[{{Kato} {et~al.}(2014){Kato}, {Saio}, {Hachisu}, \& {Nomoto}}]{Kato_2014}
{Kato}, M., {Saio}, H., {Hachisu}, I., \& {Nomoto}, K. 2014, \apj, 793, 136, \dodoi{10.1088/0004-637X/793/2/136}

\bibitem[{{Kiyota} {et~al.}(2004){Kiyota}, {Kato}, \& {Yamaoka}}]{kiyota_2004}
{Kiyota}, S., {Kato}, T., \& {Yamaoka}, H. 2004, \pasj, 56, S193, \dodoi{10.1093/pasj/56.sp1.S193}

\bibitem[{{Kumar} {et~al.}(2022a){Kumar}, {Bhalerao}, {Anupama}, {Barway}, {Basu}, {Deshmukh}, {De}, {Dutta}, {Fremling}, {Iyer}, {Jassani}, {Joharle}, {Karambelkar}, {Khandagale}, {Krishna}, {Kulkarni}, {Mate}, {Patil}, {Phanindra}, {Samantaray}, {Sharma}, {Sharma}, {Shenoy}, {Singh}, {Srivastava}, {Swain}, {Waratkar}, {Angchuk}, {Dorjay}, {Dorjai}, {Gyalson}, {Jorphail}, {Mahay}, {Norbu}, {Sharma}, {Stanzin}, {Stanzin}, \& {Stanzin}}]{harsh_2022a}
{Kumar}, H., {Bhalerao}, V., {Anupama}, G.~C., {et~al.} 2022a, \aj, 164, 90, \dodoi{10.3847/1538-3881/ac7bea}

\bibitem[{Kumar {et~al.}(2022b)Kumar, Bhalerao, Anupama, Barway, Coughlin, De, Deshmukh, Dutta, Goldstein, Jassani, Joharle, Karambelker, Khandagale, Kumar, Saraogi, Sharma, Shenoy, singer, Singh, \& Waratkar}]{harsh_2022b}
Kumar, H., Bhalerao, V., Anupama, G.~C., {et~al.} 2022b, Monthly Notices of the Royal Astronomical Society, 516, 4517, \dodoi{10.1093/mnras/stac2516}

\bibitem[{{Kumar} {et~al.}(2023){Kumar}, {Swain}, {Basu}, {Sharma}, {Barway}, {Anupama}, {Bhalerao}, \& {Angail}}]{ravi_2023}
{Kumar}, R., {Swain}, V., {Basu}, J., {et~al.} 2023, Transient Name Server AstroNote, 255, 1

\bibitem[{{Li} {et~al.}(2020){Li}, {Hambsch}, {Munari}, {Metzger}, {Chomiuk}, {Frigo}, \& {Strader}}]{Li_2020}
{Li}, K.-L., {Hambsch}, F.-J., {Munari}, U., {et~al.} 2020, \apj, 905, 114, \dodoi{10.3847/1538-4357/abc3be}

\bibitem[{{Li} {et~al.}(2017){Li}, {Metzger}, {Chomiuk}, {Vurm}, {Strader}, {Finzell}, {Beloborodov}, {Nelson}, {Shappee}, {Kochanek}, {Prieto}, {Kafka}, {Holoien}, {Thompson}, {Luckas}, \& {Itoh}}]{Li_2017}
{Li}, K.-L., {Metzger}, B.~D., {Chomiuk}, L., {et~al.} 2017, Nature Astronomy, 1, 697, \dodoi{10.1038/s41550-017-0222-1}

\bibitem[{{Lunt}(1926)}]{Lunt_1926}
{Lunt}, J. 1926, \mnras, 86, 498, \dodoi{10.1093/mnras/86.7.498}

\bibitem[{{Lyman}(2021)}]{spalipy}
{Lyman}, J.~D. 2021, {spalipy: Detection-based astronomical image registration}, Astrophysics Source Code Library, record ascl:2103.003

\bibitem[{{Mandigo-Stoba} {et~al.}(2022){Mandigo-Stoba}, {Fremling}, \& {Kasliwal}}]{MandigoStoba_2022}
{Mandigo-Stoba}, M.~S., {Fremling}, C., \& {Kasliwal}, M. 2022, The Journal of Open Source Software, 7, 3612, \dodoi{10.21105/joss.03612}

\bibitem[{{Martin} {et~al.}(2018){Martin}, {Dubus}, {Jean}, {Tatischeff}, \& {Dosne}}]{Martin_2018}
{Martin}, P., {Dubus}, G., {Jean}, P., {Tatischeff}, V., \& {Dosne}, C. 2018, \aap, 612, A38, \dodoi{10.1051/0004-6361/201731692}

\bibitem[{{Masci} {et~al.}(2019){Masci}, {Laher}, {Rusholme}, {Shupe}, {Groom}, {Surace}, {Jackson}, {Monkewitz}, {Beck}, {Flynn}, {Terek}, {Landry}, {Hacopians}, {Desai}, {Howell}, {Brooke}, {Imel}, {Wachter}, {Ye}, {Lin}, {Cenko}, {Cunningham}, {Rebbapragada}, {Bue}, {Miller}, {Mahabal}, {Bellm}, {Patterson}, {Juri{\'c}}, {Golkhou}, {Ofek}, {Walters}, {Graham}, {Kasliwal}, {Dekany}, {Kupfer}, {Burdge}, {Cannella}, {Barlow}, {Van Sistine}, {Giomi}, {Fremling}, {Blagorodnova}, {Levitan}, {Riddle}, {Smith}, {Helou}, {Prince}, \& {Kulkarni}}]{ztffp}
{Masci}, F.~J., {Laher}, R.~R., {Rusholme}, B., {et~al.} 2019, \pasp, 131, 018003, \dodoi{10.1088/1538-3873/aae8ac}

\bibitem[{Masci {et~al.}(2023)Masci, Laher, Rusholme, Shupe, Paladini, Groom, Wold, Miller, \& Drake}]{masci2023newforcedphotometryservice}
Masci, F.~J., Laher, R.~R., Rusholme, B., {et~al.} 2023, A New Forced Photometry Service for the Zwicky Transient Facility.
\newblock \doarXiv{2305.16279}

\bibitem[{{Mason} {et~al.}(2020){Mason}, {Shore}, {Kuin}, \& {Bohlsen}}]{Mason_2020}
{Mason}, E., {Shore}, S.~N., {Kuin}, P., \& {Bohlsen}, T. 2020, \aap, 635, A115, \dodoi{10.1051/0004-6361/201937025}

\bibitem[{McKinney {et~al.}(2010)}]{mckinney10}
McKinney, W., {et~al.} 2010, in Proceedings of the 9th Python in Science Conference, Vol. 445, Austin, TX, 51--56

\bibitem[{{Metzger} {et~al.}(2014){Metzger}, {Hasco{\"e}t}, {Vurm}, {Beloborodov}, {Chomiuk}, {Sokoloski}, \& {Nelson}}]{Metzger_2014}
{Metzger}, B.~D., {Hasco{\"e}t}, R., {Vurm}, I., {et~al.} 2014, \mnras, 442, 713, \dodoi{10.1093/mnras/stu844}

\bibitem[{{Mroz} \& {Udalski}(2021)}]{Mroz_2021}
{Mroz}, P., \& {Udalski}, A. 2021, The Astronomer's Telegram, 14410, 1

\bibitem[{{Mr{\'o}z} {et~al.}(2015){Mr{\'o}z}, {Udalski}, {Poleski}, {Soszy{\'n}ski}, {Szyma{\'n}ski}, {Pietrzy{\'n}ski}, {Wyrzykowski}, {Ulaczyk}, {Koz{\l}owski}, {Pietrukowicz}, \& {Skowron}}]{Mroz_2015}
{Mr{\'o}z}, P., {Udalski}, A., {Poleski}, R., {et~al.} 2015, \apjs, 219, 26, \dodoi{10.1088/0067-0049/219/2/26}

\bibitem[{{Oke} \& {Gunn}(1982)}]{Oke_1982}
{Oke}, J.~B., \& {Gunn}, J.~E. 1982, \pasp, 94, 586, \dodoi{10.1086/131027}

\bibitem[{{Oke} {et~al.}(1995){Oke}, {Cohen}, {Carr}, {Cromer}, {Dingizian}, {Harris}, {Labrecque}, {Lucinio}, {Schaal}, {Epps}, \& {Miller}}]{Oke_1995}
{Oke}, J.~B., {Cohen}, J.~G., {Carr}, M., {et~al.} 1995, \pasp, 107, 375, \dodoi{10.1086/133562}

\bibitem[{{Pandey} {et~al.}(2022){Pandey}, {Habtie}, {Bandyopadhyay}, {Das}, {Teyssier}, \& {Guarro Fl{\'o}}}]{Pandey_2022}
{Pandey}, R., {Habtie}, G.~R., {Bandyopadhyay}, R., {et~al.} 2022, \mnras, 515, 4655, \dodoi{10.1093/mnras/stac2079}

\bibitem[{{Pavana}(2020)}]{Pavana_2020}
{Pavana}, M. 2020, PhD thesis, Indian Institute of Astrophysics, Bangalore

\bibitem[{{Pavana} {et~al.}(2019){Pavana}, {Anche}, {Anupama}, {Ramaprakash}, \& {Selvakumar}}]{Pav19}
{Pavana}, M., {Anche}, R.~M., {Anupama}, G.~C., {Ramaprakash}, A.~N., \& {Selvakumar}, G. 2019, \aap, 622, A126, \dodoi{10.1051/0004-6361/201833728}

\bibitem[{{Pejcha}(2009)}]{Pejcha_2009}
{Pejcha}, O. 2009, \apjl, 701, L119, \dodoi{10.1088/0004-637X/701/2/L119}

\bibitem[{{Perley}(2019)}]{Perley_2019}
{Perley}, D.~A. 2019, \pasp, 131, 084503, \dodoi{10.1088/1538-3873/ab215d}

\bibitem[{{Pietsch}(2010)}]{Pietsch_2010}
{Pietsch}, W. 2010, Astronomische Nachrichten, 331, 187, \dodoi{10.1002/asna.200911324}

\bibitem[{{Poggiani}(2008{\natexlab{a}})}]{Poggiani_2008}
{Poggiani}, R. 2008{\natexlab{a}}, \apss, 315, 79, \dodoi{10.1007/s10509-008-9796-5}

\bibitem[{{Poggiani}(2008{\natexlab{b}})}]{Poggiani_2008a}
---. 2008{\natexlab{b}}, \na, 13, 557, \dodoi{10.1016/j.newast.2008.03.001}

\bibitem[{{Rector} {et~al.}(2022){Rector}, {Shafter}, {Burris}, {Walentosky}, {Viafore}, {Strom}, {Cool}, {Sola}, {Crayton}, {Pilachowski}, {Jacoby}, {Corbett}, {Rene}, \& {Hernandez}}]{Rector_2022}
{Rector}, T.~A., {Shafter}, A.~W., {Burris}, W.~A., {et~al.} 2022, \apj, 936, 117, \dodoi{10.3847/1538-4357/ac87ad}

\bibitem[{{Retter}(2002{\natexlab{a}})}]{Retter_2002}
{Retter}, A. 2002{\natexlab{a}}, in American Institute of Physics Conference Series, Vol. 637, Classical Nova Explosions, ed. M.~{Hernanz} \& J.~{Jos{\'e}} (AIP), 279--283, \dodoi{10.1063/1.1518214}

\bibitem[{{Retter}(2002{\natexlab{b}})}]{Retter_2002a}
{Retter}, A. 2002{\natexlab{b}}, in Astronomical Society of the Pacific Conference Series, Vol. 261, The Physics of Cataclysmic Variables and Related Objects, ed. B.~T. {G{\"a}nsicke}, K.~{Beuermann}, \& K.~{Reinsch}, 655, \dodoi{10.48550/arXiv.astro-ph/0110102}

\bibitem[{{Ryon} \& {Stark}(2023)}]{Ryon_2023}
{Ryon}, J.~E., \& {Stark}, D.~V. 2023, in ACS Instrument Handbook for Cycle 32 v. 23.0, Vol.~23, 23

\bibitem[{{Schaefer}(2021)}]{Schaefer_2021}
{Schaefer}, B.~E. 2021, Research Notes of the American Astronomical Society, 5, 150, \dodoi{10.3847/2515-5172/ac0d5b}

\bibitem[{{Schwarz} {et~al.}(1997){Schwarz}, {Hauschildt}, {Starrfield}, {Baron}, {Allard}, {Shore}, \& {Sonneborn}}]{Schwarz_1997}
{Schwarz}, G.~J., {Hauschildt}, P.~H., {Starrfield}, S., {et~al.} 1997, \mnras, 284, 669, \dodoi{10.1093/mnras/284.3.669}

\bibitem[{{Schwarz} {et~al.}(1998){Schwarz}, {Hauschildt}, {Starrfield}, {Whitelock}, {Baron}, \& {Sonneborn}}]{Schwarz_1998}
---. 1998, \mnras, 300, 931, \dodoi{10.1046/j.1365-8711.1998.01964.x}

\bibitem[{{Schwarz} {et~al.}(2001){Schwarz}, {Shore}, {Starrfield}, {Hauschildt}, {Della Valle}, \& {Baron}}]{Schwarz_2001}
{Schwarz}, G.~J., {Shore}, S.~N., {Starrfield}, S., {et~al.} 2001, \mnras, 320, 103, \dodoi{10.1046/j.1365-8711.2001.03960.x}

\bibitem[{{Schwarz} {et~al.}(2007){Schwarz}, {Woodward}, {Bode}, {Evans}, {Eyres}, {Geballe}, {Gehrz}, {Greenhouse}, {Helton}, {Liller}, {Lyke}, {Lynch}, {O'Brien}, {Rudy}, {Russell}, {Shore}, {Starrfield}, {Temim}, {Truran}, {Venturini}, {Wagner}, {Williams}, \& {Zamanov}}]{Schwarz_2007}
{Schwarz}, G.~J., {Woodward}, C.~E., {Bode}, M.~F., {et~al.} 2007, \aj, 134, 516, \dodoi{10.1086/519240}

\bibitem[{{Shafter} {et~al.}(2011){Shafter}, {Bode}, {Darnley}, {Misselt}, {Rubin}, \& {Hornoch}}]{Shafter_2011}
{Shafter}, A.~W., {Bode}, M.~F., {Darnley}, M.~J., {et~al.} 2011, \apj, 727, 50, \dodoi{10.1088/0004-637X/727/1/50}

\bibitem[{{Shafter} {et~al.}(2009){Shafter}, {Rau}, {Quimby}, {Kasliwal}, {Bode}, {Darnley}, \& {Misselt}}]{Shafter_2009}
{Shafter}, A.~W., {Rau}, A., {Quimby}, R.~M., {et~al.} 2009, \apj, 690, 1148, \dodoi{10.1088/0004-637X/690/2/1148}

\bibitem[{{Shafter} {et~al.}(2015){Shafter}, {Henze}, {Rector}, {Schweizer}, {Hornoch}, {Orio}, {Pietsch}, {Darnley}, {Williams}, {Bode}, \& {Bryan}}]{Shafter_2015}
{Shafter}, A.~W., {Henze}, M., {Rector}, T.~A., {et~al.} 2015, \apjs, 216, 34, \dodoi{10.1088/0067-0049/216/2/34}

\bibitem[{{Shafter} {et~al.}(2024){Shafter}, {Hornoch}, {Ku{\v{c}}{\'a}kov{\'a}}, {Fatka}, {Zhao}, {Gao}, {Yaqup}, {Zhong}, {Esamdin}, {Bai}, {Wang}, {Benni}, {Luo}, \& {Yousuf}}]{Shafter_2024}
{Shafter}, A.~W., {Hornoch}, K., {Ku{\v{c}}{\'a}kov{\'a}}, H., {et~al.} 2024, Research Notes of the American Astronomical Society, 8, 5, \dodoi{10.3847/2515-5172/ad19de}

\bibitem[{{Shara} {et~al.}(2018){Shara}, {Prialnik}, {Hillman}, \& {Kovetz}}]{Shara_2018}
{Shara}, M.~M., {Prialnik}, D., {Hillman}, Y., \& {Kovetz}, A. 2018, \apj, 860, 110, \dodoi{10.3847/1538-4357/aabfbd}

\bibitem[{{Singh} {et~al.}(2014){Singh}, {Tandon}, {Agrawal}, {Antia}, {Manchanda}, {Yadav}, {Seetha}, {Ramadevi}, {Rao}, {Bhattacharya}, {Paul}, {Sreekumar}, {Bhattacharyya}, {Stewart}, {Hutchings}, {Annapurni}, {Ghosh}, {Murthy}, {Pati}, {Rao}, {Stalin}, {Girish}, {Sankarasubramanian}, {Vadawale}, {Bhalerao}, {Dewangan}, {Dedhia}, {Hingar}, {Katoch}, {Kothare}, {Mirza}, {Mukerjee}, {Shah}, {Shah}, {Mohan}, {Sangal}, {Nagabhusana}, {Sriram}, {Malkar}, {Sreekumar}, {Abbey}, {Hansford}, {Beardmore}, {Sharma}, {Murthy}, {Kulkarni}, {Meena}, {Babu}, \& {Postma}}]{KPS_2014}
{Singh}, K.~P., {Tandon}, S.~N., {Agrawal}, P.~C., {et~al.} 2014, in Society of Photo-Optical Instrumentation Engineers (SPIE) Conference Series, Vol. 9144, Space Telescopes and Instrumentation 2014: Ultraviolet to Gamma Ray, ed. T.~{Takahashi}, J.-W.~A. {den Herder}, \& M.~{Bautz}, 91441S, \dodoi{10.1117/12.2062667}

\bibitem[{{Sirianni} {et~al.}(2005){Sirianni}, {Jee}, {Ben{\'\i}tez}, {Blakeslee}, {Martel}, {Meurer}, {Clampin}, {De Marchi}, {Ford}, {Gilliland}, {Hartig}, {Illingworth}, {Mack}, \& {McCann}}]{Sirianni_2005}
{Sirianni}, M., {Jee}, M.~J., {Ben{\'\i}tez}, N., {et~al.} 2005, \pasp, 117, 1049, \dodoi{10.1086/444553}

\bibitem[{{Sokolovsky} {et~al.}(2020){Sokolovsky}, {Aydi}, {Chomiuk}, {Kawash}, {Strader}, {Mukai}, {Li}, {Page}, {Beardmore}, \& {Osborne}}]{Sokolovsky_2020}
{Sokolovsky}, K., {Aydi}, E., {Chomiuk}, L., {et~al.} 2020, The Astronomer's Telegram, 14078, 1

\bibitem[{Stanek \& Garnavich(1998)}]{Stanek_1998}
Stanek, K.~Z., \& Garnavich, P.~M. 1998, The Astrophysical Journal, 503, L131, \dodoi{10.1086/311539}

\bibitem[{Starrfield(1999)}]{STARRFIELD1999371}
Starrfield, S. 1999, Physics Reports, 311, 371, \dodoi{https://doi.org/10.1016/S0370-1573(98)00116-1}

\bibitem[{Starrfield {et~al.}(2016)Starrfield, Iliadis, \& Hix}]{Starrfield_2016}
Starrfield, S., Iliadis, C., \& Hix, W.~R. 2016, Publications of the Astronomical Society of the Pacific, 128, 051001, \dodoi{10.1088/1538-3873/128/963/051001}

\bibitem[{{Starrfield} {et~al.}(2012){Starrfield}, {Timmes}, {Iliadis}, {Hix}, {Arnett}, {Meakin}, \& {Sparks}}]{Starrfield_2012}
{Starrfield}, S., {Timmes}, F.~X., {Iliadis}, C., {et~al.} 2012, Baltic Astronomy, 21, 76, \dodoi{10.1515/astro-2017-0361}

\bibitem[{{Steinberg} \& {Metzger}(2020)}]{Steinberg_2020}
{Steinberg}, E., \& {Metzger}, B.~D. 2020, \mnras, 491, 4232, \dodoi{10.1093/mnras/stz3300}

\bibitem[{{Strope} {et~al.}(2010){Strope}, {Schaefer}, \& {Henden}}]{strope_2010}
{Strope}, R.~J., {Schaefer}, B.~E., \& {Henden}, A.~A. 2010, \aj, 140, 34, \dodoi{10.1088/0004-6256/140/1/34}

\bibitem[{{Tanaka} {et~al.}(2011{\natexlab{a}}){Tanaka}, {Nogami}, {Fujii}, {Ayani}, \& {Kato}}]{Tanaka_2011}
{Tanaka}, J., {Nogami}, D., {Fujii}, M., {Ayani}, K., \& {Kato}, T. 2011{\natexlab{a}}, \pasj, 63, 159, \dodoi{10.1093/pasj/63.1.159}

\bibitem[{{Tanaka} {et~al.}(2011{\natexlab{b}}){Tanaka}, {Nogami}, {Fujii}, {Ayani}, {Kato}, {Maehara}, {Kiyota}, \& {Nakajima}}]{Tanaka_2011a}
{Tanaka}, J., {Nogami}, D., {Fujii}, M., {et~al.} 2011{\natexlab{b}}, \pasj, 63, 911, \dodoi{10.1093/pasj/63.4.911}

\bibitem[{{Tandon} {et~al.}(2020){Tandon}, {Postma}, {Joseph}, {Devaraj}, {Subramaniam}, {Barve}, {George}, {Ghosh}, {Girish}, {Hutchings}, {Kamath}, {Kathiravan}, {Kumar}, {Lancelot}, {Leahy}, {Mahesh}, {Mohan}, {Nagabhushana}, {Pati}, {Rao}, {Sankarasubramanian}, {Sriram}, \& {Stalin}}]{Tandon_2020}
{Tandon}, S.~N., {Postma}, J., {Joseph}, P., {et~al.} 2020, \aj, 159, 158, \dodoi{10.3847/1538-3881/ab72a3}

\bibitem[{{Tody}(1993)}]{tod93}
{Tody}, D. 1993, in Astronomical Society of the Pacific Conference Series, Vol.~52, Astronomical Data Analysis Software and Systems II, ed. R.~J. {Hanisch}, R.~J.~V. {Brissenden}, \& J.~{Barnes}, 173

\bibitem[{{Vacca} {et~al.}(2003){Vacca}, {Cushing}, \& {Rayner}}]{Vacca_2003}
{Vacca}, W.~D., {Cushing}, M.~C., \& {Rayner}, J.~T. 2003, \pasp, 115, 389, \dodoi{10.1086/346193}

\bibitem[{Van~Rossum \& Drake(2009)}]{python09}
Van~Rossum, G., \& Drake, F.~L. 2009, Python 3 Reference Manual (Scotts Valley, CA: CreateSpace)

\bibitem[{Virtanen {et~al.}(2020)Virtanen, Gommers, Oliphant, Haberland, Reddy, Cournapeau, Burovski, Peterson, Weckesser, Bright, {van der Walt}, Brett, Wilson, Millman, Mayorov, Nelson, Jones, Kern, Larson, Carey, Polat, Feng, Moore, {VanderPlas}, Laxalde, Perktold, Cimrman, Henriksen, Quintero, Harris, Archibald, Ribeiro, Pedregosa, {van Mulbregt}, \& {SciPy 1.0 Contributors}}]{SciPy20}
Virtanen, P., Gommers, R., Oliphant, T.~E., {et~al.} 2020, Nature Methods, 17, 261, \dodoi{10.1038/s41592-019-0686-2}

\bibitem[{{Warner}(2003)}]{Warner_2003}
{Warner}, B. 2003, {Cataclysmic Variable Stars}, \dodoi{10.1017/CBO9780511586491}

\bibitem[{{Warner}(2008)}]{Warner_2008}
---. 2008, in Classical Novae, ed. M.~F. {Bode} \& A.~{Evans}, Vol.~43, 16--33, \dodoi{10.1017/CBO9780511536168.004}

\bibitem[{{Williams}(2012)}]{Williams_2012}
{Williams}, R. 2012, \aj, 144, 98, \dodoi{10.1088/0004-6256/144/4/98}

\bibitem[{{Williams} {et~al.}(2013){Williams}, {Bode}, {Darnley}, {Evans}, {Zubko}, \& {Shafter}}]{Williams_2013}
{Williams}, S.~C., {Bode}, M.~F., {Darnley}, M.~J., {et~al.} 2013, \apjl, 777, L32, \dodoi{10.1088/2041-8205/777/2/L32}

\bibitem[{{Wilson} {et~al.}(2004){Wilson}, {Henderson}, {Herter}, {Matthews}, {Skrutskie}, {Adams}, {Moon}, {Smith}, {Gautier}, {Ressler}, {Soifer}, {Lin}, {Howard}, {LaMarr}, {Stolberg}, \& {Zink}}]{Wilson_2004}
{Wilson}, J.~C., {Henderson}, C.~P., {Herter}, T.~L., {et~al.} 2004, in Society of Photo-Optical Instrumentation Engineers (SPIE) Conference Series, Vol. 5492, Ground-based Instrumentation for Astronomy, ed. A.~F.~M. {Moorwood} \& M.~{Iye}, 1295--1305, \dodoi{10.1117/12.550925}

\bibitem[{{Yaron} {et~al.}(2005){Yaron}, {Prialnik}, {Shara}, \& {Kovetz}}]{Yaron_2005}
{Yaron}, O., {Prialnik}, D., {Shara}, M.~M., \& {Kovetz}, A. 2005, \apj, 623, 398, \dodoi{10.1086/428435}

\end{thebibliography}
\bibliographystyle{aasjournal}

\end{document}